\documentclass[lettersize,journal]{IEEEtran}
\IEEEoverridecommandlockouts
\usepackage{amsmath,amsfonts}
\usepackage{amssymb}
\usepackage[caption=false,font=footnotesize,labelfont=sf,textfont=sf]{subfig}
\usepackage{stfloats}
\usepackage{url}
\usepackage{graphicx}
\usepackage{cite}
\usepackage{float}
\usepackage{subfig}
\usepackage{multirow}
\usepackage{color}
\usepackage{makecell}
\usepackage{booktabs}
\usepackage[colorlinks,citecolor=blue,linkcolor=blue]{hyperref}
\newtheorem{corollary}{Corollary}
\newtheorem{example}{Example}

\newtheorem{theorem}{Theorem}

\newtheorem{property}{Property}
\newtheorem{definition}{Definition}
\newtheorem{algorithm}{Algorithm}
\usepackage{algorithm, algorithmicx, algpseudocode}
\begin{document}

\title{Polarization Decomposition and Its Applications}

\author{Tianfu Qi, \emph{Graduate Student Member, IEEE}, Jun Wang, \emph{Senior Member, IEEE}
\thanks{Tianfu Qi, Jun Wang are with the National Key Laboratory on Wireless Communications, University of Electronic Science and Technology of China, Chengdu 611731, China (e-mail: 202311220634@std.uestc.edu.cn).}
}

\maketitle

\begin{abstract}
The polarization decomposition of arbitrary binary-input memoryless channels (BMCs) is studied in this work. By introducing the polarization factor (PF), defined in terms of the conditional entropy of the channel output under various input configurations, we demonstrate that the symmetric capacities of the polarized subchannels can be uniformly expressed as functions of the PF. The explicit formulation of the PF as a function of the block length and subchannel index is derived. Furthermore, an efficient algorithm is proposed for the computation of the PF. Notably, we establish a one-to-one correspondence between each PF and an $n$-ary tree. Leveraging this tree structure, we develop a pruning method to determine the conditional entropy associated with different input relationships. The proposed polarization framework offers both theoretical insights and practical advantages, including intuitive visualization of polarization behavior and efficient polar code construction. To the best of our knowledge, this is the first approach that enables the efficient computation of symmetric capacities for all subchannels in arbitrary BMCs.
\end{abstract}


\section{Introduction}
As a significant breakthrough in coding theory, polar codes are the first class that can be theoretically proven to achieve the capacity of binary-input discrete memoryless channels (B-DMCs) as the block length tends to infinity \cite{paper1}. These codes are constructed based on the principle of channel polarization, which is realized through recursive processes of channel combining and channel splitting. As the polarization process progresses, the synthesized subchannels asymptotically evolve into either noiseless or completely noisy channels, with their capacities approaching 1 and 0, respectively. Furthermore, the fraction of noiseless subchannels among the total number of synthesized subchannels converges to the capacity of the underlying channel, denoted by $I(W)$.

The mutual information (MI) process $\{I_n, 1 \leq n \leq \log L\}$\footnote{In this paper, $\log$ denotes the base-2 logarithm.} and the Bhattacharyya parameter process $\{Z_n, 1 \leq n \leq \log L\}$ are two fundamental components underlying the polarization phenomenon, where $L$ represents the block length. The properties of the process $\{Z_n\}$ have been extensively studied in the literature \cite{paper1,paper2,paper3,paper4,paper5,paper7,paper10}. For example, \cite{paper10} analyzes the polarization rate by examining the convergence behavior of $\{Z_n\}$, which in turn is used to derive an upper bound on the decoding error probability of polar codes. The asymptotic behavior of channel polarization and the error exponent of polar codes as functions of block length are investigated in \cite{paper2} and \cite{paper5}, respectively. Moreover, the Bhattacharyya parameter serves as a key metric in polar code construction, as it effectively characterizes channel reliability. Subchannels with higher error probabilities are typically assigned frozen bits, which are predetermined values known to both the encoder and decoder. In \cite{paper7}, an approximation method for the evolution of ${Z_n}$ applicable to arbitrary B-DMCs is proposed, which enables the design of a practical encoding algorithm.

In contrast to the Bhattacharyya parameter process $\{Z_n\}$, the MI process $\{I_n\}$ has received relatively limited attention, primarily due to its computational intractability. Notably, recursive computation of MI is only feasible for the binary erasure channel (BEC) case \cite{paper1}. For general binary-input memoryless channels (BMCs), the computational complexity of evaluating $\{I_n\}$ increases rapidly with block length, thereby limiting its practical utility. Nevertheless, accurate estimation of symmetric capacity remains highly valuable. For instance, a fundamental goal of polar code construction is to identify and select polarized subchannels with higher capacities. This task becomes straightforward if the MI of all subchannels is available.

Another critical issue in polar coding is rate loss, which arises from insufficient polarization at finite block lengths. Although polar codes are capable of achieving the channel capacity asymptotically, this ideal is unattainable in practical settings. For example, \cite{paper5} shows that the gap between the achievable rate and the channel capacity scales as $L^{-1/\mu}$, where $\mu = 3.627$ for BECs and $\mu < 4.714$ for any B-DMCs. In this work, we aim to precisely quantify the rate loss for a given block length and channel condition with low computational complexity. These challenges can be effectively addressed if the symmetric capacities of the subchannels are accurately and efficiently computed.

Based on the definition of MI for polarized subchannels, it can be observed that the channel output exhibits partial independence from the input source. Specifically, the MI of the $i$-th subchannel can be expressed as $I(Y_1^L, U_1^{i-1}; U_i)$, wherein certain components of $Y_1^L$ are statistically independent of $U_i$. This structural property allows for a further decomposition of the conventional MI expression. In this work, \textbf{our objective is to reformulate the MI of polarized subchannels into a more intuitive and computationally tractable form applicable to arbitrary BMCs.} Motivated by the above observations, the main contributions of this paper are summarized as follows,
\begin{itemize}
\item{We first introduce a general framework that characterizes the representational relationships among the bits within a polar codeword. Then, we define the polarization factor (PF), a real-valued metric that captures the dependency structure between two sets of channel inputs. The PF quantitatively equals the conditional entropy of channel outputs with corresponding channel input relations.}
\item{Leveraging the recursive structure of the polar code generation matrix, we explicitly derive analytical expressions for the PF corresponding to arbitrary block lengths and subchannel indices. Furthermore, we demonstrate that the conventional MI of subchannels can be represented as a combination of PFs, offering a more tractable and analytically convenient form for both theoretical investigation and practical computation.}
\item{An efficient algorithm is further developed to compute the PF. In particular, we show that each PF can be uniquely associated with an $n$-ary tree structure, in which nodes are categorized into two distinct types. By designing pruning operations for each node type, the computation of the PF is reduced to evaluating the entropy of channel outputs under various input power configurations. To facilitate further applications, we also derive closed-form expressions for the channel output entropy.}
\item{We demonstrate that the overall computational complexity of the polarization decomposition algorithm is $\mathcal{O}(L^{\log 3} \log L)$. The proposed algorithm is amenable to parallel implementation due to the inherent tree structure.}
\item{The applications of the proposed polarization decomposition are twofold. From a theoretical perspective, the transformed expressions of subchannel mutual information enable the analysis of relationships among subchannels, including the verification of partial orders (PO) in polar codes. From a practical standpoint, the proposed decomposition algorithm allows for the efficient computation of symmetric capacities across all subchannels with acceptable complexity. This facilitates the design of MI-based polar code construction schemes and enables direct estimation of rate loss under finite block lengths.}
\end{itemize}

The remainder of this paper is organized as follows. In section \ref{section_2}, we briefly revisit the fundamentals of polar codes and introduce the key definitions used throughout the paper. Section \ref{section_3} presents the transformation of subchannel mutual information into a combination of PFs, along with the derivation of their explicit expressions. In section \ref{section_4}, we construct the $n$-ary tree representation of PFs and propose an efficient algorithm for their computation. Building on these results, the polarization decomposition algorithm is developed and its theoretical complexity is analyzed. The theoretical and practical applications of polarization decomposition are discussed in section \ref{section_5}. Finally, section \ref{section_6} concludes the paper and outlines potential directions for future research.

\emph{Notations:} Uppercase letters denote random variables (RVs), while lowercase letters represent their realizations, e.g., $X$ and $x$. A vector is denoted as $X_a^b = [X_a, \cdots, X_b]$ with $X_a^b = \emptyset$ if $b < a$. To avoid ambiguity when representing multiple vectors, we also use bold notation $\vec{X}$ to denote a vector and $\vec{X}_a^b$ to denote $(b - a + 1)$ vectors. $\oplus$ denotes modulo-2 addition, and $\bigoplus_{i=1}^{N} X_i \triangleq X_1 \oplus \cdots \oplus X_N$. $\otimes$ represents the Kronecker product and $F^{\otimes a}$ denotes the $a$-fold Kronecker product of matrix $F$. The cardinality of a set $\mathcal{A}$ is written as $|\mathcal{A}|$ and $P^\top$ represents the transpose of matrix $P$. `$\mathbf{1}$' is a constant vector or matrix of ones with appropriate dimension subscripts. The ceiling and floor functions are denoted by $\lceil \cdot \rceil$ and $\lfloor \cdot \rfloor$, respectively.

\section{Preliminary}\label{section_2}
\subsection{Polar code}
Let the block length be $L=2^l$. Define the function $W:\mathcal{X}\rightarrow\mathcal{Y}$ and $W_L^{(i)}:\mathcal{U}\rightarrow\mathcal{U}^{i-1}\times\mathcal{Y}^L$ to represent the original channel and the polarized subchannel, respectively. Here, we restrict that $\mathcal{U},\mathcal{X}\in\mathcal{B}=\{0,1\}$ and $W$ belongs to the BMCs. In the following, we mainly focus on the case that $\mathcal{Y}$ is continuous-valued. As the block length approaches infinity, the symmetric capacity $I(W_L^{(i)})$ converges to 0 or 1 almost surely. Let $\mathcal{A}(L)$ denote the set of indices corresponding to the noiseless subchannels for block length $L$. Then, we have $\lim\limits_{L \rightarrow +\infty} |\mathcal{A}(L)| = I(W)$. The primary task in polar code construction is thus to identify the information set $\mathcal{A}(L)$.

Denote the uncoded source vector and coded vector separately as $u_1^L$ and $x_1^L$. These two vectors are related through the generator matrix $B_L$. The $B_L$ can be constructed by the $l$-fold Kronecker product of polarization kernel, i.e.,
\begin{equation}\label{B_L}
B_L=F^{\otimes l}
\end{equation}
and
$F=\left[
\begin{array}{cc}
  1 & 0 \\
  1 & 1
\end{array}
\right]$. Then, we have $x_1^L=u_1^LB_LR_L$ where $R_L$ is the bit-reversal permutation matrix \cite{paper1}.

\textbf{Remark 1:} Without loss of generality, we omit the influence of $R_L$ in the subsequent analysis since it only reorders the subchannels and does not affect the mutual information of each polarized subchannel. For the sake of simplicity, we use the expression $x_1^L = u_1^L B_L$ throughout the remainder of this paper.

\subsection{Basic definitions}
In this subsection, several foundational concepts are introduced to support the subsequent analysis. To clarify the definitions and avoid potential ambiguities, illustrative examples are also provided.
\begin{definition}\label{definition_1}
\textit{Let $Y_1^p$ and $\tilde{Y}_1^p$ be two channel output random vectors corresponding to the input random vectors $X_1^p$ and $\tilde{X}_1^p$, respectively. Then, the serial combination (SC) entropy and parallel combination (PC) entropy are defined as follows,
\begin{equation}\label{SC_definition}
\text{SC: }h_S(p)\triangleq h(Y|Y_1^p),
\end{equation}
\begin{equation}\label{PC_definition}
\text{PC: }h_P(p)\triangleq h(\tilde{Y}|\tilde{Y}_1^p),
\end{equation}
where $X=X_1\oplus X_2\oplus\cdots\oplus X_p$ and $\tilde{X}=\tilde{X}_1=\cdots=\tilde{X}_p$.}
\end{definition}

It is evident that $h_S(1) = h_P(1)$. Moreover, both sequences ${h_S(p)}$ and ${h_P(p)}$ are monotonic with respect to $p$. The sequence ${h_P(p)}$ can be interpreted as transmitting the bit $\tilde{x}$ for $p$ times. As a result, the corresponding conditional entropy $h_P(p)$ decreases as $p$ increases. A similar analysis applies to the sequence ${h_S(p)}$. Consequently, the inequalities $h_S(1) \leq h_S(2) \leq \cdots \leq h_S(+\infty)$ and $h_P(1) \geq h_P(2) \geq \cdots \geq h_P(+\infty)$ hold.

Following this, we discuss the asymptotic and monotonic properties of the SC entropy and PC entropy.

\begin{property}\label{property_1}
\textit{Define the $h_S(p)$ and $h_P(p)$ as in definition \ref{definition_1}. Denote the detection error probability of $x$ given $y$ by $P_e$ and $0<P_e\leq1/2$. Then,
\begin{equation}\label{limit_SC_entropy_order}
\lim\limits_{p\rightarrow+\infty}h_S(p)=h(Y)
\end{equation}
\begin{equation}\label{limit_PC_entropy_order}
\lim\limits_{p\rightarrow+\infty}h_P(p)=h(N)
\end{equation}
where $h(N)$ is the channel noise entropy.}
\end{property}
\begin{IEEEproof}
The proof is relegated to appendix \ref{appendix_property_1}.
\end{IEEEproof}

The property \ref{property_1} is consistent with the preceding intuitive analysis of monotonicity. It should be noted that the expression \eqref{limit_SC_entropy_order} does not hold when the channel is perfectly noiseless. In this case, the value of $x$ is fully determined by any observation $y_j$. As the signal-to-noise ratio (SNR) approaches infinity, it follows that $h_S(p) = h_P(p) = h(N)$ for all $p$.

\begin{property}\label{property_2}
\textit{Define the $h_S(p)$ and $h_P(p)$ as in definition \ref{definition_1}. Assume that the power of the corresponding channel inputs $X_j$ and $\tilde{X}_j$ is $P_0$ for all $j = 1, \cdots, p$. Then,
\begin{equation}\label{limit_SC_entropy_power_large}
\lim\limits_{P_0\rightarrow+\infty}h_S(p)=h(N)
\end{equation}
\begin{equation}\label{limit_SC_entropy_power_small}
\lim\limits_{P_0\rightarrow0}h_S(p)=h(N)
\end{equation}
\begin{equation}\label{limit_PC_entropy_power_large}
\lim\limits_{P_0\rightarrow+\infty}h_P(p)=h(N)
\end{equation}
\begin{equation}\label{limit_PC_entropy_power_small}
\lim\limits_{P_0\rightarrow0}h_P(p)=h(N)
\end{equation}
for any fixed $p$.}
\end{property}
\begin{IEEEproof}
The proof is straightforward, and we focus on the case of $h_S(p)$ since the same reasoning applies to $h_P(p)$. The case of infinite SNR has already been addressed. When $P_0 \rightarrow 0$, it becomes impossible to extract any information about $x_1^p$ from $y_1^p$. In this completely noisy scenario, there is no correlation between $y_1^p$ and $y$. Consequently, we have
$\lim\limits_{P_0 \rightarrow 0} h_S(p) = \lim\limits_{P_0 \rightarrow 0} h(Y) = h(N)$.
\end{IEEEproof}

\begin{figure}[htbp]
\centering
\includegraphics[width=3.5in]{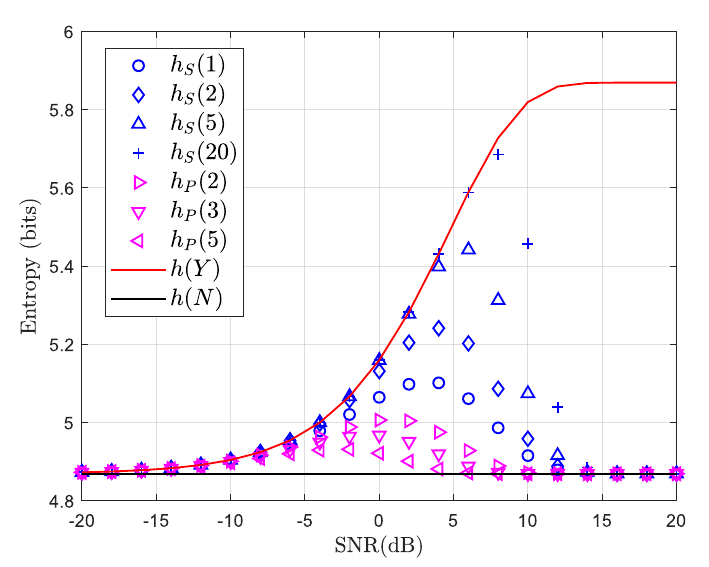}
\caption{The comparison of SC entropy and PC entropy for various values of $p$ is presented. The considered channel noise is white Gaussian noise (WGN). The channel output entropy $h(Y)$ and noise entropy $h(N)$ are also provided for comparison. It can be observed that both $h_S(p)$ and $h_P(p)$ are equal to $h(N)$ for both small and sufficiently large SNR. Moreover, the relationship
$h(N) = h_P(+\infty) \leq \cdots \leq h_P(1) \leq h_S(1) \leq \cdots \leq h_S(+\infty) = h(Y)$
holds for all SNR values. The $h_S(p)$ is quite close to $h(Y)$ in the low SNR regime, but it decreases to $h(N)$ as the SNR continues to increase. Furthermore, the decay rate is much faster for larger values of $p$ compared to smaller ones.}
\label{fig_1}
\end{figure}

Intuitively, the entropy is expected to increase monotonically with respect to the channel input power. However, Property \ref{property_2} reveals that for any fixed value of $p$, neither $h_S(p)$ nor $h_P(p)$ is monotonic with respect to the SNR. In particular, both quantities initially increase as the SNR rises, but begin to decrease once the SNR exceeds a certain threshold. An illustrative example is provided in Fig. \ref{fig_1} to further validate this behavior. The accurate and efficient method for computing these quantities will be detailed in section \ref{section_4}.

\begin{definition}\label{definition_2}
\textit{Let block length be $L$ and the $i$-th polarization submatrix be $\tilde{B}_L(i)=[\mathbf{0}_{L\times(i-1)},B_L(i:L,:)^\top]^\top$. Then, the $i$-th subcode $\mathcal{C}_L(i)$ is defined as $\mathcal{C}_L(i)=\{x_1^L:x_1^L=u_1^L\tilde{B}_L(i),u_1^L\in\mathcal{U}^L\}$. The codeword and output vector corresponding to the $i$-th subcode are denoted as $x_1^L(\mathcal{C}_L(i))$ and $y_1^L(\mathcal{C}_L(i))$, respectively. Their RV version is $X_1^L(\mathcal{C}_L(i))$ and $Y_1^L(\mathcal{C}_L(i))$.}
\end{definition}

We denote the element at the $j$-th row and $k$-th column of $\tilde{B}_L(i)$ by $\tilde{B}_L(i, [j, k])$. Clearly, we have $B_L = \tilde{B}_L(1)$. Unlike $B_L$, the matrix $\tilde{B}_L(i)$ for $i > 1$ is not invertible, as the elements of the first $(i-1)$ rows are set to zero. Consequently, the column rank of $\tilde{B}_L(i)$ is strictly less than $L$ for $1 < i \leq L$. For any codeword $x_1^L(\mathcal{C}_L(i))$ in $\mathcal{C}_L(i)$, some bits may be expressed as linear combinations of other bits. For example, we have
\begin{align}
x_j(\mathcal{C}_L(i))=&u_1^L\tilde{B}_L(i,[:,j])\nonumber\\
=&\bigoplus_{k=i}^{L}u_k\tilde{B}_L(i,[k,j])
\end{align}
where elements of $\tilde{B}_L(i)$ in the previous $(i-1)$ rows are zeros. Therefore, expressing $x_j(\mathcal{C}_L(i))$ in terms of $x_{j+1}^L(\mathcal{C}_L(i))$ is equivalent to representing the column vector $\tilde{B}_L(i, [:, j])$ using $\tilde{B}_L(i, [:, j+1:L])$. This equivalence is crucial for deriving an explicit expression for the simplified subchannel MI.

Next, we provide operations to explicitly express the relationship between $x$ and $x_1^p$, which are different bits of a codeword in $\mathcal{C}_L(i)$.

\begin{definition}\label{definition_3}
\textit{The representation relation between different bits, $x$ and $x_1^p$, of a codeword in $\mathcal{C}_L(i)$ is denoted by $R(x, x_1^p)$ and determined by the following three operations.
\begin{enumerate}
\item{SC operation: If $x=\bigoplus_{i=1}^{p}x_i$ and $x_i,i=1,\cdots,p$ are mutually different, we define $R(x,x_1^p)=[(p)^1_S]$. Furthermore, if there are several distinct representations, i.e., $x=\bigoplus_{i=1+(j-1)p}^{jp}x_i$, $j=1,\cdots,J$ and $x_i,i=1,\cdots,Jp$ are mutually different, we define $R(x,x_1^p)=[(p)^J_S]$.}
\item{PC operation: If $x=x_i,i=1,\cdots,p$, we define $R(x,x_1^p)=[(1)^p_P]$.}
\item{Overlap operation: If $x=x_0\oplus(\bigoplus_{i=1+(j-1)p}^{jp}x_i)$, $j=1,\cdots,J$ and $x_i,i=1,\cdots,Jp$ are mutually distinct, we define $R(x,x_1^p)=[(p)^J_S\leftarrow 1]$. On the other hand, if $x_{i}=x_{i+p}=\cdots=x_{i+(J-1)p},i=1,\cdots,p$ but for a fixed $j=1,\cdots,J$, $\{x_{1+(j-1)p},\cdots,x_{p+(j-1)p}\}$ are mutually diverse, we define $R(x,x_1^p)=[(p)^J_P\leftarrow 1]$.}
\end{enumerate}}
\end{definition}

\textbf{Remark 2:} The relation expression is composed of the factor $(a)^b_{\chi}$. Here, $a$ and $b$ denote the element number in a representation and the number of representations, respectively. $\chi \in {S, P}$ is the combination operation indicator, which indicates whether the elements between different representations are mutually independent or identical.

\textbf{Remark 3:} The three basic operations can be further concatenated and nested if the relation becomes more complex. For instance, the $a_1$ in $(a_1)^{b_1}_{\chi_1}$ can be replaced by $(a_2)^{b_2}_{\chi_2}$. Furthermore, the argument in the `overlap operation' can also be substituted by SC operation, PC operation, or more complicated structures, e.g., $(a_1)^{b_1}_{\chi_1}\leftarrow(a_2)^{b_2}_{\chi_1}$.

\textbf{Remark 4:} For $a_1$ in $(a_1)^{b_1}_{\chi_1}$, we omit its indicator since there is no ambiguity based on the value of $a_1$. Indeed, the complete expression should be $({a_1}^{b_0}_{\chi_0})$. When $a_1=1$ and $b_0=1$, $(1^1_S)$ and $(1^1_P)$ are the same and written as $(1)$. If $a_1=1$ and $b_0>1$, the combination operation must be PC, and we have $(1^{b_0}_K)$. If $a_1>1$ and $b_0=1$, it indicates that the representation contains $a_1$ different elements, which cannot be identical and must follow the PC operation. Otherwise, $({a_1}^1_P)$ could be converted to $(1^{a_1}_P)$. Finally, $({a_1}^{b_0}_{\chi_0})$ with $a_1>1,b_0>1$ can be rewritten as $(a_1)^{b_0}_{\chi_0}$, which coincides to the original form $(a_1)^{b_1}_{\chi_1}$.

\textbf{Remark 5:} The factor can be simplified for specific values of $a$ and $b$. When $b=1$, $(a)^b_{\chi}$ is equivalent to $(a)$. Additionally, if $a=0$ or $b=0$, the term $(a)^b_{\chi}$ simplifies to $(0)$ and can be omitted .

\textbf{Remark 6:} Adjacent identical operations can be merged. For example, we have $((a_1)^{b_1}_{S})^{b_2}_{S}=(a_1)^{b_1b_2}_{S}$ and $((a_1)^{b_1}_{P})^{b_2}_{P}=(a_1)^{b_1b_2}_{P}$. However, $(((a_1)^{b_1}_{S})^{b_2}_{P})^{b_3}_{S}$ cannot be further simplified.

To avoid confusion, we provide an example to further explain the above definitions and analysis.
\begin{example}\label{example_1}
\textit{Let the block length $L=8$. We aim to examine the relationship between the first bit of the codeword $x_1^8(\mathcal{C}_8(i)),i=1,\cdots,8$ and other bits. Based on $\tilde{B}_8(i)$, it can be observed that
\begin{align}
R(x_1(\mathcal{C}_8(1)),x_2^8(\mathcal{C}_8(1)))&=[(0)]\\
R(x_1(\mathcal{C}_8(2)),x_2^8(\mathcal{C}_8(2)))&=[(7)]\\
R(x_1(\mathcal{C}_8(3)),x_2^8(\mathcal{C}_8(3)))&=[(3)]\label{example_2_3}\\
R(x_1(\mathcal{C}_8(4)),x_2^8(\mathcal{C}_8(4)))&=[(3)^3_S\leftarrow1]\label{example_2_4}\\
R(x_1(\mathcal{C}_8(5)),x_2^8(\mathcal{C}_8(5)))&=[(1)]\\
R(x_1(\mathcal{C}_8(6)),x_2^8(\mathcal{C}_8(6)))&=[(1),(3)^2_P]\label{example_2_6}\\
R(x_1(\mathcal{C}_8(7)),x_2^8(\mathcal{C}_8(7)))&=[(1)^3_P]\\
R(x_1(\mathcal{C}_8(8)),x_2^8(\mathcal{C}_8(8)))&=[(1)^7_P]
\end{align}}

\textit{Note that for $i=1$, we have $\tilde{B}_8(1)=B_8$ and $\tilde{B}_8(1)$ has full column rank. Thus, $x_1(\mathcal{C}_8(1))$ is different from $x_2^8(\mathcal{C}_8(1))$. Next, we explain \eqref{example_2_3} as an example, it can be verified that $x_1(\mathcal{C}_8(3))=x_3(\mathcal{C}_8(3))\oplus x_5(\mathcal{C}_8(3))\oplus x_7(\mathcal{C}_8(3))$. Similar analyses can be applied to other cases.}
\end{example}

Now, we are ready to define the PF, which can be utilized to transform conventional MI expressions and is more general than SC entropy and PC entropy. The motivation behind introducing the PF is that the functions $h_S(\cdot)$ and $h_P(\cdot)$ are insufficient to accurately express the subchannel MI as $L$ increases. According to example \ref{example_1}, both \eqref{example_2_4} and \eqref{example_2_6} cannot be expressed by SC and PC entropy.

\begin{definition}\label{definition_4}
\textit{Suppose we aim to represent $x$ by $x_1^{p}$ where $x$ and $x_j$, $j=1,\cdots,p$ are distinct bits of the codeword $x_1^L(\mathcal{C}_L(i))$. We define the polarization factor $\lambda(R(x, x_1^{p}))$ as a mapping from the relation between $x$ and $x_1^{p}$ to the corresponding conditional entropy, i.e.,
\begin{align}\label{polarization_factor}
\lambda(R(x,x_1^{p}))\triangleq h(Y|Y_1^{p})
\end{align}
where $Y$ and $Y_1^p$ separately corresponds to the channel output of $X$ and $X_1^p$.}
\end{definition}

The PF is general and can be reduced to SC entropy and PC entropy in specific cases. For instance, we have $\lambda([1^p_P]) = h_P(p)$ and $\lambda([p]) = h_S(p)$. With the explicit expression of $R(x, x_1^{p})$, the conditional entropy of $Y$ given $Y_1^p$ can be easily determined. To further clarify it, we provide an example.

\begin{example}\label{example_2}
\textit{Consider the calculation of $\lambda([1^p_P])$. Based on definitions \ref{definition_3} and \ref{definition_4}, we have $\lambda([1^p_P]) = h_P(p)= h(Y, Y_1^p) - h(Y_1^p)$ and $X = X_i$ for $i = 1, \cdots, p$. Denote $h(Y) = h_Y(d = A)$, where the corresponding input RV $X \in \{0, A\}$ with $A$ being a constant. Thus, by applying the coordinate transformation, we obtain
\begin{align}
\lambda([1^p_P])=&h_Y(d=\sqrt{p+1}A)+h(N)\nonumber\\
&-h_Y(d=\sqrt{p}A)
\end{align}
where $h(N)$ is the noise entropy. In this case, the remaining task is to derive the analytical expression for $h_Y(d = A)$, which is significantly easier than directly computing high-dimensional integration.}
\end{example}

\section{Polarized subchannel decomposition}\label{section_3}
In this section, we first decompose the MI expressions of polarized subchannels into a combination of PFs. Subsequently, we derive closed-form expressions for the PFs applicable to arbitrary block lengths and subchannel indices. Owing to the complexity of the derivation, we categorize the analysis into several distinct cases.

\subsection{Transformation of $I(W_L^{(i)})$}
The MI of a polarized subchannel is generally difficult to compute through numerical integration. However, the term $I(Y_1^L, U_1^{i-1}; U_i)$ can be further simplified, since $U_1^{i-1}$ serves as prior information that reduces the uncertainty of $X_1^L$ corresponding to $Y_1^L$. In addition, some components of $Y_1^L$ are independent of $U_1^i$ and can be omitted. Based on the chain rule and the definition of subcodes,
\begin{align}\label{transform of MI_1}
I(W_L^{(i)})=&I(Y_1^L;U_i|U_1^{i-1})\nonumber\\
=&I(Y_1^L(\mathcal{C}_L(i));U_i)\nonumber\\
=&I(Y_{\mathcal{D}_i}(\mathcal{C}_L(i));U_i|Y_{\mathcal{D}^c_i}(\mathcal{C}_L(i)))
\end{align}
where $\mathcal{D}_i$ denotes the set of indices corresponding to those components of $Y_j$ that are related to $U_i$. Equivalently, $\mathcal{D}_i$ corresponds to the index set in which the elements of the $i$-th row of $\tilde{B}_L(i)$ are nonzero. For example, when $L=4$, we have $\mathcal{D}_1 = \{1\}$, $\mathcal{D}_2 = \{1, 2\}$, $\mathcal{D}_3 = \{1, 3\}$, and $\mathcal{D}_4 = \{1, 2, 3, 4\}$. \textbf{Throughout this paper, we explicitly omit the permutation matrix for simplicity.} Let $\mathcal{D}_i(j)$ denote the $j$-th element of $\mathcal{D}_i$, where $j = 1, \cdots, |\mathcal{D}_i|$. Thus,
\begin{align}\label{transform of MI_2}
&I(W_L^{(i)})=\sum_{j=1}^{|\mathcal{D}_i|}h(Y_{\mathcal{D}_i(j)}(\mathcal{C}_L(i))|Y_{\{\mathcal{D}^c_i,\mathcal{D}_i(1:j-1)\}}(\mathcal{C}_L(i)))\nonumber\\
&+h(Y_{\mathcal{D}_i(j)}(\mathcal{C}_L(i+1))|Y_{\{\mathcal{D}^c_i,\mathcal{D}_i(1:j-1)\}}(\mathcal{C}_L(i+1)))
\end{align}

Then, the subchannel MI can be decomposed by the PF as follows,
\begin{align}\label{transform of MI_3}
&I(W_L^{(i)})=\sum_{j=1}^{|\mathcal{D}_i|}\lambda(R(x_{\mathcal{D}_i(j)}(\mathcal{C}_L(i))|x_{\{\mathcal{D}^c_i,\mathcal{D}_i(1:j-1)\}}(\mathcal{C}_L(i))))\nonumber\\
&+\lambda(R(x_{\mathcal{D}_i(j)}(\mathcal{C}_L(i+1))|x_{\{\mathcal{D}^c_i,\mathcal{D}_i(1:j-1)\}}(\mathcal{C}_L(i+1))))
\end{align}

\subsection{Explicit expressions of PF}
The \eqref{transform of MI_3} remains implicit and thus cannot be directly utilized, as the underlying relationships vary with the subchannel index and the block length. To address this, we proceed to derive analytical expressions for PFs. The derivation is based on the recursive structure of the generator matrix.
\begin{equation}\label{iterative_structure}
B_L=B_{L/2}\otimes F
\end{equation}

However, deriving a unified expression that accommodates arbitrary subchannel indices and block lengths is highly complex. Therefore, we divide the problem into several manageable cases. Specifically, from \eqref{transform of MI_3}, the two terms in the summation correspond to two distinct scenarios: one involves computing the joint entropy of $Y_1^L(\mathcal{C}_L(i))$ without the prior knowledge of $U_i$, and the other includes this prior information. 

We then separately consider the four cases $i\bmod4=j,j=0,1,2,3$. In fact, it seems more natural to consider odd and even subchannel indices due to the polarization kernel. However, we observe that the cases $i \bmod 4 = 0$ and $i \bmod 4 = 2$ exhibit subtle differences that prevent them from being represented using a single unified expression. Combining these cases would introduce unnecessary complexity.

Furthermore, we distinguish between the index ranges $1 \leq i \leq L/2$ and $L/2 + 1 \leq i \leq L$. This separation is motivated by the fact that for $L/2 + 1 \leq i \leq L$, the first $L/2$ rows of $\tilde{B}_L(i)$ are identical to its last $L/2$ columns. This repetition leads to a different structural relationship that must be handled separately. In total, we derive 16 distinct expressions for the polarization function. Nevertheless, we will demonstrate that these expressions are not entirely independent. Instead, they exhibit structural similarities and can be transformed into one another through appropriate modifications.

Before proceeding, we introduce several parameters to facilitate the subsequent analysis.
\begin{enumerate}
\item{The Hamming weight of the $i$-th row of $\tilde{B}_L(i)$ is defined by $H_L(i),i=1,\cdots,L$.}
\item{The repetition number of columns of $\tilde{B}_L(i)$ is denoted by $\beta_L(i),i=1,\cdots,L$. In other words, we have that $\tilde{B}_L(i,[:,(j-1)L/\beta_L(i)+1:jL/\beta_L(i)]),j=1,\cdots,\beta_L(i)$ are the same.}
\item{The first consecutive 1's number of the $i$-th row of $\tilde{B}_L(i)$ is denoted by $C_L(i),i=1,\cdots,L$.}
\item{The maximum number of element among all representations of $\tilde{B}_L(i,[:,1])$ by $\tilde{B}_L(i,[:,2:L])$ is denoted by $\hat{\theta}_L(i),i=2,\cdots,L$. Furthermore, we define $\theta_L(i)\triangleq\log(\hat{\theta}_L(i)+1)$.}
\item{The minimum number of element among all representations of $\tilde{B}_L(i,[:,1])$ by $\tilde{B}_L(i,[:,2:L])$ is denoted by $\hat{\varepsilon}_L(i),i=2,\cdots,L$. Furthermore, we define $\varepsilon_L(i)\triangleq\log(\hat{\varepsilon}_L(i)+1)$.}
\end{enumerate}

It is important to note that for $i = 1$, the matrix $\tilde{B}_L(1)$ has full column rank, and therefore the parameters $\hat{\theta}_L(1)$ and $\hat{\varepsilon}_L(1)$ are not defined. Additionally, with regard to $\hat{\theta}_L(i)$, the maximum count pertains only to the number of elements involved in a single representation. For instance, consider the case where $x = x_j \oplus x_{j+1}$ for $j = 1, 2, 3$. Although it is mathematically true that $x = \bigoplus_{j=1}^6 x_j$, the correct value of $\hat{\theta}_L(i)$ should be 2 instead of 6, as each valid representation involves only two variables. We now present a concrete example to illustrate the definitions of these parameters.

\begin{example}
\textit{Let block length $L=8$. The values of the five aforementioned parameters are summarized in Table \ref{table_1}.
\begin{table}[htbp]
\begin{center}
\caption{Parameter values for $L=8$}\label{table_1}
\begin{tabular}{p{0.2cm}<{\centering}p{0.55cm}<{\centering}p{0.55cm}<{\centering}p{0.55cm}<{\centering}p{0.8cm}<{\centering}p{0.8cm}<{\centering}p{0.55cm}<{\centering}p{0.55cm}<{\centering}}
\toprule
$i$&$H_L(i)$&$\beta_L(i)$&$C_L(i)$&$\hat{\theta}_L(\lceil \frac{i}{4}\rceil)$&$\theta_L(\lceil \frac{i}{4}\rceil)$&$\hat{\varepsilon}_L(i)$&$\varepsilon_L(i)$\\
\midrule
1& 1& 1& 1& $\times$& $\times$& $\times$& $\times$\\
2& 2& 1& 2& 7& 3& 7& 3\\
3& 2& 1& 1& 3& 2& 3& 2\\
4& 4& 1& 4& 3& 2& 3& 2\\
5& 2& 2& 1& 1& 1& 1& 1\\
6& 4& 2& 2& 3& 2& 1& 1\\
7& 4& 4& 1& 1& 1& 1& 1\\
8& 8& 8& 8& 1& 1& 1& 1\\
\bottomrule
\end{tabular}
\end{center}
\end{table}}
\end{example}

It is worth noting that all five parameters can be determined through iterative procedures. The corresponding computation methods are provided below.

\begin{equation}\label{iterative_H}
H_L(i)=2H_L(i-2^{\lceil\log i\rceil-1}),
\end{equation}
with $H_L(1)=1$.

\begin{equation}\label{iterative_beta}
\beta_L(i)=\frac{L}{2^{\lceil\log(L+1-i)\rceil}}.
\end{equation}

\begin{equation}
C_L(i)=\left\{
\begin{aligned}\label{iterative_C}
&i,i=2^2,2^3,\cdots,L\\
&K(i\bmod4+1),\text{others}
\end{aligned}
\right.
\end{equation}
with $K(1:4)=(4,1,2,1)$.

\begin{align}\label{iterative_theta}
\theta_L(L+2-&2^{i+1}:L)=[\times,\theta_L(L+2-2^{i}:L)+1,1,\nonumber\\
&\theta_L(L+2-2^{i}:L)],i=1,\cdots,\log L-1
\end{align}
with $\theta_L(1)=\times$, $\theta_L(L)=1$ and $i=1,\cdots,\log L-1$.

\begin{equation}\label{iterative_varepsilon}
\varepsilon_L(i)=\log L+1-\lceil\log i\rceil.
\end{equation}

Here, we explain the expression of $\theta_L(i)$ and other parameter expressions can be easily derived. The expression $\theta_L(L+1-2^{i}) = 1$ in \eqref{iterative_theta} holds because $\tilde{B}_L(L+1-2^k)$ consists of $\frac{L}{2^k}$ identical $2^k \times 2^k$ submatrices, all of which have full column rank. Therefore, $\tilde{B}_L(L+1-2^k,[:,1])$ can be expressed in terms of $\tilde{B}_L(2^k,[:,1+j\frac{L}{2^k}])$ for $k = 0, \dots, \log L-1$ and $j = 1, \dots, \frac{L}{2^k} - 1$.

For the expression $\theta_L(2:L-2^{i}) = \theta_L(L+2-2^{i}:L) + 1$, we use the case $L = 4$ as an example to facilitate understanding. In this case, we observe that $\hat{\theta}_4(2) = 3$. The procedure is as follows: To represent $\tilde{B}_4(2,[:,1])$, we must select $\tilde{B}_4(2,[:,2])$ because $\tilde{B}_4(2,[2,2])$ equals $\tilde{B}_4(2,[2,1])$. Then, for $\tilde{B}_4(2,[3:4,1])$, it is equivalent to $\tilde{B}_4(2,[3:4,3])$ due to the properties of the Kronecker product. However, the influence of $\tilde{B}_4(2,[3:4,2])$ should be excluded. Therefore, the column $\tilde{B}_4(2,[:,4])$ needs to be selected. \textbf{We refer to this process as the `padding-removing' step.} In summary, after one padding-removing' step, the number of representation elements $\hat{\theta}$ increases according to the relation $\hat{\theta} \to 2\hat{\theta} + 1$, which is equivalent to $\theta \to \theta + 1$.

We now present the first theorem which focuses on the expression of $R(x_{\mathcal{D}_i(j)}(\mathcal{C}_L(i))|x_{\{\mathcal{D}^c_i,\mathcal{D}_i(1:j-1)\}}(\mathcal{C}_L(i)))$ with $i\bmod4=0$ and $1\leq i\leq L/2$.

\begin{theorem}\label{theorem_1}
\textit{Let the block length be $L$ and assume the subchannel index satisfies $i\bmod 4 = 0$. Define a constant sequence $\{Q_0,\cdots,Q_M\}$ and an index sequence $\{q_0,\cdots,q_M\}$, where $q_j\in\{1,\cdots,Q_j\},j=0,\cdots,M$. Let
\begin{equation}\label{theorem_1_t_j}
t_j\triangleq t_{j-1}\bmod(2^{\lceil\log t_{j-1}\rceil}/\beta_{2^{\lceil\log t_{j-1}\rceil}}(t_{j-1}))
\end{equation}
with $t_0=i$ and $j=1,\cdots,M-1$. The term $Q_j$ is defined as
\begin{align}\label{theorem_1_Q}
Q_{M-j+1}=\min\Big\{\frac{H_L(i)}{C_L(i)\prod_{h=1}^{j-1}Q_{M-h+1}},&\beta_{2^{\lceil\log t_{j-1}\rceil}}(t_{j-1})\Big\},\nonumber\\
&\hspace{-0.2cm} j=1,\cdots,M
\end{align}
and $Q_0=C_L(i)$. The integer $M$ is the largest index such that $Q_{j},j=0,\cdots,M$ are all larger than 1. Define
\begin{equation}\label{theorem_1_T_k}
T_k\triangleq2^{\theta_L(i)-k}-1
\end{equation}
\begin{align}\label{theorem_1_phi}
\phi_k(q_{M-k+1}^M)\triangleq&[(\phi_{k-1}(q_{M-k+2}^M),\nonumber\\
&((T_{M-k+3})^{\prod_{h=1}^{k-1}Q_{M+1-h}}_S))^{q_{M-k+1}}_S],\nonumber\\
&\hspace{3cm} k=2,\cdots,M
\end{align}
with $\phi_1(q_M)=(T_{M+1})^{q_M}_S$. Define column vector sequences $\{E_h^-\},\{E_h^+\}$ for $h=1,\cdots,M$, initialized as $E_1^-=(q_{M})^\top$ and $E_1^-=(q_{M}-1)^\top$. The vectors $\{E_h^-\},\{E_h^+\},h=2,\cdots,M$ can be obtained iteratively,
\begin{align}\label{theorem_1_E_h_minus}
E_h^-=&((E_{h-1}^-,q_{M-h+1}\mathbf{1}_{2^{h-2}})^\top,\nonumber\\
&(E_{h-1}^+,(Q_{M-h+1}-q_{M-h+1})\mathbf{1}_{2^{h-2}})^\top)^\top
\end{align}
\begin{align}\label{theorem_1_E_h_plus}
E_h^+=&((E_{h-1}^-,(q_{M-h+1}-1)\mathbf{1}_{2^{h-2}})^\top,\nonumber\\
&(E_{h-1}^+,(Q_{M-h+1}-q_{M-h+1}+1)\mathbf{1}_{2^{h-2}})^\top)^\top
\end{align}
where $\mathbf{1}_{a}$ denotes a column vector of ones with dimension $a$. $(E_{h-1}^-,q_{M-h+1}\mathbf{1}_{2^{h-2}})$ represents their concatenation. Then,
\begin{align}\label{theorem_1_omega_minus}
\Omega_k^-=(\phi_{M-k}(E_{M-k}^-(1)),\cdots,\phi_{M-k}(E_{M-k}^-(2^{M-k-1})))
\end{align}
\begin{align}\label{theorem_1_omega_plus}
\Omega_k^+=(\phi_{M-k}(E_{M-k}^+(1)),\cdots,\phi_{M-k}(E_{M-k}^+(2^{M-k-1})))
\end{align}}

\textit{Finally, for $1\leq i\leq L/2$,
\begin{align}\label{theorem_1_final_equation}
&\sum_{j=1}^{|\mathcal{D}_i|}\lambda(R(x_{\mathcal{D}_i(j)}(\mathcal{C}_L(i))|x_{\{\mathcal{D}^c_i,\mathcal{D}_i(1:j-1)\}}(\mathcal{C}_L(i))))\nonumber\\
=&\sum_{q_M=1}^{Q_M}\cdots\sum_{q_1=1}^{Q_1}\sum_{q_0=1}^{C_L(i)}\lambda([(2^\epsilon_L(i)-1)^{q_M-1}_S,\Theta_{M-1}]\nonumber\\
&\hspace{4cm} \leftarrow(2^{\epsilon_L(i)-1}-1)),
\end{align}
where for $k=0,\cdots,M-1$, we define
\begin{equation}\label{theorem_1_Theta}
\Theta_{k}\triangleq((V_{k}^-,V_{k}^+,\Theta_{k-1})\leftarrow (T_{k+1}-T_{k+2})_S^{\prod_{h=k+1}^{M}Q_h})
\end{equation}
\begin{equation}\label{theorem_1_V_k_minus}
V_k^-=((T_k)^{\prod_{h=k+1}^{M}Q_h}_S\leftarrow\Omega_k^-)^{q_k-1}_S
\end{equation}
\begin{equation}\label{theorem_1_V_k_plus}
V_k^+=((T_k)^{\prod_{h=k+1}^{M}Q_h}_S\leftarrow\Omega_k^+)^{Q_k-q_k}_S
\end{equation}
with $\Theta_{-1}=(0)$.}
\end{theorem}

\begin{figure}[htbp]
\centering
\includegraphics[width=3.6in]{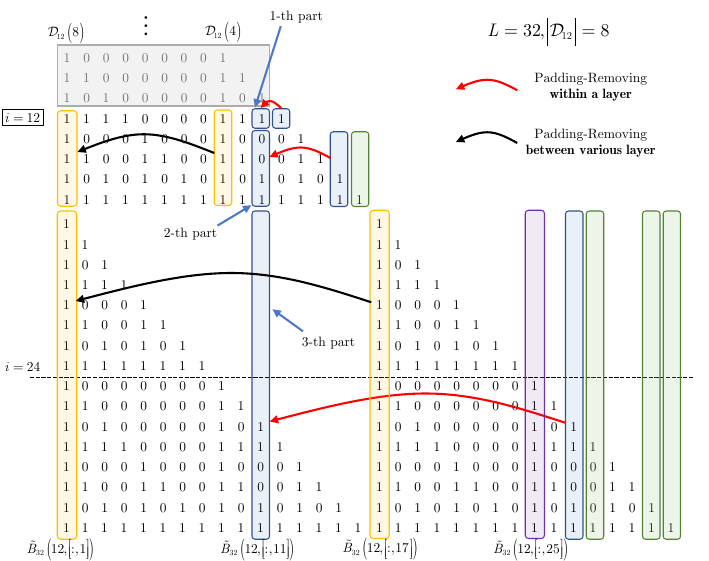}
\caption{We set $L = 32$ and $i = 12$, and the figure presents the submatrix of $\tilde{B}_{32}(12)$ due to space limitations. The missing parts of the matrix are all zeros. Additionally, the gray section represents bits that have been considered as prior information, and thus their influence can be omitted. An example of representations within a layer and across different layers is provided. Specifically, we aim to represent $\tilde{B}_{32}(12,[:,11])$ using columns that contain $\tilde{B}_{32}(12,[:,12])$, and to represent $\tilde{B}_{32}(12,[:,1])$ using columns that contain $\tilde{B}_{32}(12,[:,9])$. Elements within the blue and green boxes correspond to representations within a single layer, while elements across different layers are marked by yellow and purple boxes.}
\label{fig_2}
\end{figure}

\begin{figure*}[htbp]
\centering
\subfloat[Case 1]{\includegraphics[width=3.6in]{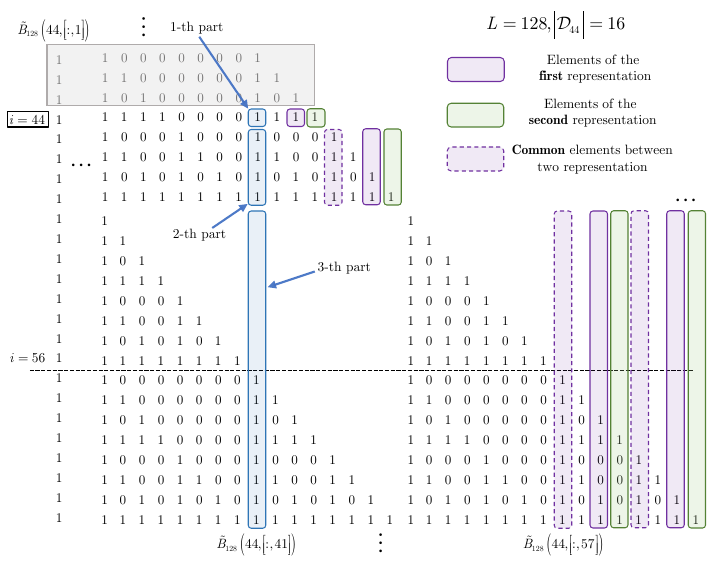}\label{fig_2_plus_1}}
\subfloat[Case 2]{\includegraphics[width=3.6in]{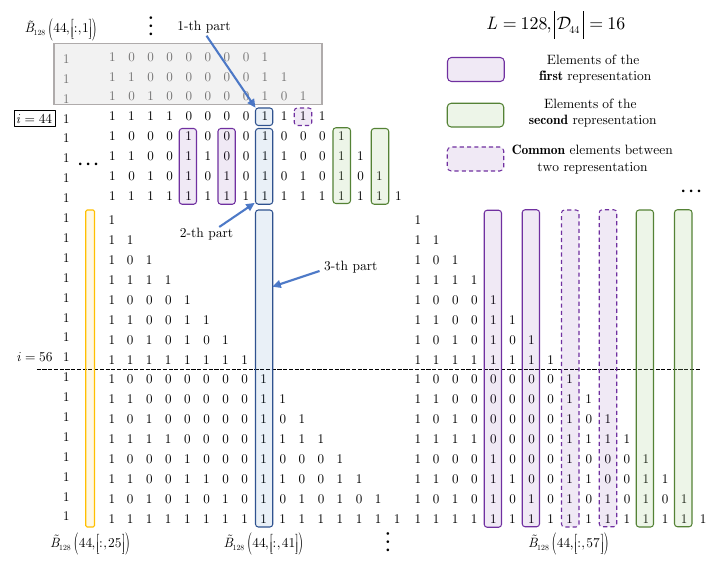}\label{fig_2_plus_2}}
\caption{We set $L = 128$ and $i = 44$, and the figure shows the submatrix of $\tilde{B}_{128}(44)$. An example of representations both within a layer and across different layers is provided. For each case, we consider two representations for $\tilde{B}_{128}(44,[:,41])$. The purple dashed boxes highlight the common elements shared by the two representations. Note that we only present three `padding-removing' steps due to space limitations.}
\label{fig_2_plus}
\end{figure*}

\begin{IEEEproof}
The proof of theorem \ref{theorem_1} is somewhat intricate, and to clarify the argument, we will divide it into several steps. \textbf{The complexity arises primarily from the fact that for a given target column, there may be multiple representations, with some elements being shared between different representations. The purpose of this theorem is to fully characterize the underlying rules governing these representations.}

\textbf{Step 1: Motivation for dividing $|\mathcal{D}_i|$ into several layers.}

The number of elements in each representation may vary depending on the index $j$ in $\mathcal{D}_i(j)$. To address this, we partition the index set $\mathcal{D}_i$ into multiple layers, where each layer contains representations with the same number of elements and admits a uniform relation form. Note that we treat each sequence of $C_L(i)$ consecutive `1's in the $i$-th row of $\tilde{B}_L(i)$ as the innermost layer, with $q_0$ serving as its index indicator. Thus, we have $q_0 \in {1, \dots, C_L(i)}$. The representation form for each 1' within the consecutive 1's remains consistent, as the submatrix $\tilde{B}_L(i,[i+1:i+C_L(i), i_D - C_L(i) + 1 : i_D])$ is a replica of $\tilde{B}_L(i,[i+1:i+C_L(i), i_D+1 : i_D + C_L(i)])$, where $i_D$ is the column index corresponding to $\mathcal{D}_i(1)$. Moreover, these two submatrices have full column rank. Therefore, the $C_L(i)$ columns are linearly independent and cannot be used to represent one another. Consequently, the partition ${Q_1, \dots, Q_M}$ must satisfy $|\mathcal{D}i| = C_L(i) \prod_{j=1}^{M} Q_j$.

\textbf{Step 2: Representation relation within the innermost layer.}

We now describe the representation process in the innermost layer, as illustrated in Fig. \ref{fig_2}. We set $L = 32$ and aim to express $\tilde{B}_{32}(12,[:,11])$. We divide $\tilde{B}_{32}(12,[12:32,11])$, which needs to be constructed, into three parts: $\tilde{B}_{32}(12,[12,11])$, $\tilde{B}_{32}(12,[13:16,11])$, and $\tilde{B}_{32}(12,[17:32,11])$. We consider using $\tilde{B}_{32}(12,[:,12])$, $\tilde{B}_{32}(12,[:,15])$, and $\tilde{B}_{32}(12,[:,27])$ to pad these three parts, respectively. The columns used for padding and removing are separately marked by blue and green boxes in Fig. \ref{fig_2}.

For the first part, there is no need to remove the influence of elements from previous parts. For the second part, $\tilde{B}_{32}(12,[:,15])$ is selected to pad $\tilde{B}_{32}(12,[13:16,11])$. Meanwhile, the influence of the elements used in previous parts on $\tilde{B}_{32}(12,[13:16,11])$ must be removed. Therefore, $\tilde{B}_{32}(12,[:,16])$ is chosen to cancel the effect of $\tilde{B}_{32}(12,[13:16,12])$. A similar analysis applies to the third part. Consequently, after each step, the number of representations $\hat{\theta}$ undergoes the transformation $\hat{\theta} \rightarrow 2\hat{\theta} + 1$. The factor of $2$' comes from the removal operation, while the $+1$' accounts for padding the corresponding part. Note that the relation $\hat{\theta} \rightarrow 2\hat{\theta} + 1$ is equivalent to $\theta \rightarrow \theta + 1$. Finally, we conclude that the representation size of the innermost layer (or equivalently, the first layer) is given by $T_0$.

\textbf{Step 3: Representation relation across various layers.}

The representation across various layers differs from the representation within a single layer, and we describe this via Fig. \ref{fig_2}. The focused elements are marked by yellow and purple boxes. For $\tilde{B}_{32}(12,[:,1])$, similar to the analysis in `Step 2', it should initially be divided into three parts. However, when we apply $\tilde{B}_{32}(12,[:,9])$ to express $\tilde{B}_{32}(12,[:,1])$, we observe that $\tilde{B}_{32}(12,[12:16,1]) = \tilde{B}_{32}(12,[12:16,9])$, so the first and second parts can be merged. As a result, the number of `padding-removing' steps decreases by 1. 

When we express $\tilde{B}_{32}(12,[:,j])$ using $\tilde{B}_{32}(12,[:,j+8])$ for $j = 1, \dots, 4$, the step number is smaller than the corresponding representation within the previous layer by 1. The underlying reason is that $\tilde{B}_{32}(12,[9:16,1:8])$ is replicated from $\tilde{B}_{32}(12,[9:16,9:16])$ via the Kronecker operation. Thus, $\tilde{B}_{32}(12,[12:16,j])$ can be directly padded by $\tilde{B}_{32}(12,[12:16,j+8])$. Each time the number of `padding-removing' steps decreases by 1, we consider this representation as belonging to a higher layer.

For instance, expressing $\tilde{B}_{32}(12,[:,1])$ using columns containing $\tilde{B}_{32}(12,[:,2])$ or $\tilde{B}_{32}(12,[:,10])$ corresponds to the first layer. However, if we construct it using $\tilde{B}_{32}(12,[:,9])$ and other columns, it belongs to the second layer. Therefore, the representation size of the $k$-th layer is denoted by $T_k$.

\textbf{Step 4: Common elements between different representations.}

The presence of common elements between different representations complicates the derivation of explicit expressions. To address this, we utilize Fig. \ref{fig_2_plus} to explain the analysis process, considering two distinct cases.

First, we examine the scenario depicted in Fig. \ref{fig_2_plus_1}. In this case, the representation should be divided into four parts, but due to space constraints, we only present the first three `padding-removing' steps. The two representations for $\tilde{B}_{128}(44,[:,41])$ are marked by purple and green boxes. The padding elements for the first part are distinct. For the second part, the two representations share the common element $\tilde{B}_{128}(44,[:,45])$. At the third step, to pad $\tilde{B}_{128}(44,[49:64,41])$ and eliminate the influence of $\tilde{B}_{128}(44,[:,45])$, two additional common elements are selected. Based on this analysis, the number of common elements across different representations in the $k$-th layer equals $T_{k+1}$.

The second case, illustrated in Fig. \ref{fig_2_plus_2}, involves a fixed padding element for the first part. Here, $\tilde{B}_{128}(44,[:,43])$ is the common element, and $\tilde{B}_{128}(44,[:,59])$ is chosen to remove the influence on $\tilde{B}_{128}(44,[49:64,43])$. It is important to note that $\tilde{B}_{128}(44,[:,57])$ is also common, but it appears in both case 1 and case 2. As a result, we handle this element separately and do not count it again as the common element in case 2. In this way, the number of common elements that occur exclusively in case 2, and not in both case 1 and case 2, is given by $(T_{k+1}-T_{k+2})$. The term $T_{k+2}$ refers to the common elements shared between cases 1 and 2, while $T_{k+1}$ represents the total number of common elements across the two representations, as analyzed in case 1.

\textbf{Step 5: Analysis of $\{Q_j\}$.}

We have defined that the first layer consists of the $C_L(i)$ consecutive `1's in the $i$-th row. Thus, the sequence $\{Q_1, \cdots, Q_M\}$ satisfies
\begin{equation}\label{constraint_Q_sequence}
\prod_{j=1}^{M}Q_j=\frac{H_L(i)}{C_L(i)}
\end{equation}
where ${Q_1,\cdots,Q_M}=\emptyset$ when $M=0$. If $H_L(i)/C_L(i) = 1$, then all `1's in the $i$-th row are consecutive and thus belong to a single layer, namely the first layer. Thus, we have $M=0$. For the case $H_L(i)/C_L(i) > 1$, we begin the derivation from $Q_M$, which is associated with the repetition number of the covering matrix of the $i$-th row. This covering matrix is defined as the smallest submatrix of $B_{L}$ that contains all elements relevant to a fixed index in the considered layer. Moreover, the size of the covering matrix must be a power of two, i.e., $2^l$.

For example, consider the case $i=12$ in Fig. \ref{fig_2}. There are two layers and hence $M=1$. For the second layer, one of its covering matrices is $\tilde{B}_{32}(12,[1:8,1:8])$. This is justified as follows. The full covering matrix that includes all `1's is $\tilde{B}_{32}(12,[1:16,1:16])$. We then restrict our analysis to this submatrix. For $q_1=0$ and $q_1=1$, the corresponding element ranges are $\tilde{B}_{32}(12,[12,1:8])$ and $\tilde{B}_{32}(12,[12,9:16])$, respectively. Consequently, the covering matrices are separately $\tilde{B}_{32}(12,[1:8,1:8])$ and $\tilde{B}_{32}(12,[1:8,9:16])$.

It is observed that the number of covering matrices in the $k$-th layer can be used to be $Q_k$. The determination of ${Q_j}$ is inherently an iterative process. To formalize this, we define the operator $\mathcal{M}(\cdot)$ as
\begin{equation}
\mathcal{M}^l(i)\triangleq i\bmod\Big(\frac{L}{2^{\lceil\log \mathcal{M}^{l-1}(i)\rceil}}\Big),l\geq1
\end{equation}
where $\mathcal{M}^0(i)=i$. For the $M$-th layer, its covering matrix is $\tilde{B}_{L}(i,[1:2^{\lceil\log i\rceil},1:2^{\lceil\log i\rceil}])$ and then, $Q_M$ equals $\beta_{2^{\lceil\log i\rceil}}(i)$. For the $(M-1)$-th layer, a smaller covering matrix is considered, typically of size $2^{\lceil\log \mathcal{M}(i)\rceil}$. Here, the operator $\mathcal{M}(\cdot)$ plays a critical role in updating the row index used in the next-layer covering matrix, ensuring it remains within the valid range of the matrix. Following this iterative process, it follows that
\begin{align}
Q_{k}=\beta_{2^{\lceil\log \mathcal{M}^{k-1}(i)\rceil}}(\mathcal{M}^{k-1}(i))
\end{align}

Finally, $i=2^l,l=2,\cdots,\log L$ are special cases in which the previously described procedure for computing $Q_j$ is not directly applicable. Specifically, for these values of $i$, we have $H_L(i)/C_L(i)=1$ and thus, $M=0$ or equivalently, $Q_j$ for arbitrary $j\geq1$. To accommodate these cases, we introduce an additional constraint on the computation of $Q_j$. Namely, each $Q_j$ must satisfy  $Q_j\leq H_L(i)/(C_L(i)\prod_{h=1}^{j-1}Q_{M-h+1})$. Then, \eqref{theorem_1_Q} is established.

\textbf{Step 6: Analysis of $\Omega_k^-$ and $\Omega_k^+$.}

The sets $\Omega_k^-$ and $\Omega_k^+$ denote the common elements specific to the $k$-th layer. As established in `Step 4', both $|\Omega_k^-|$ and $|\Omega_k^+|$ are equal to $(T_{k+1} - T_{k+2})$. Referring to Fig. \ref{fig_2_plus_2}, the element $\tilde{B}_{128}(44,[:,59])$ serves as a common term for eliminating the influence of $\tilde{B}_{128}(44,[:,43])$. Similarly, $\tilde{B}_{128}(44,[:,27])$ can serve the same function, owing to the equivalence $\tilde{B}_{128}(44,[49:64,49:64]) = \tilde{B}_{128}(44,[17:32,17:32])$. For the removal of elements associated with the second part, the number of candidates equals the number of covering matrices in the first layer, which is $Q_2 Q_1$. However, we omit the corresponding notation in the final expression because, for a fixed column used to pad the first part, there exist $Q_2 Q_1$ possible representations for the target column. Thus, based on the property of relation operation, we have $[(a_1)^{b_1}_S\leftarrow(a_2)^{b_1}_S]=[(a_1)^{b_1}_S]$ where $a_1\geq a_2$. Hence, it follows that the elements responsible for padding and removal in the $k$-th layer appear a total of $\prod_{h=k}^{M} Q_h$ times.

Then, we derive the sequences ${E_h^-}$ and ${E_h^+}$ corresponding to $\Omega_k^-$ and $\Omega_k^+$, respectively. These sequences represent the number of occurrences of common elements used for removing operations across different layers. It is important to note that padding elements are not included in $\Omega_k^-$ or $\Omega_k^+$, as they are essential for every representation and therefore not specific to any particular representation. Consequently, all common elements considered here are associated with the removing operation, and their usage is determined by the column chosen for padding the first part.

To illustrate the relationships between common elements used for removal, we again refer to Fig. \ref{fig_2_plus_2}. In this scenario, we have $M = 2$, $Q_2 = Q_1 = 2$, and $Q_0 = 4$. Consider the task of expressing $\tilde{B}_{128}(44,[:,41])$ using columns that include $\tilde{B}_{128}(44,[:,43])$. Here, $\tilde{B}_{128}(44,[:,43])$ is selected to pad the first part. To remove the influence of $\tilde{B}_{128}(44,[45:48,43])$, the following columns can be used: $\tilde{B}_{128}(44,[:,7])$, $\tilde{B}_{128}(44,[:,15])$, $\tilde{B}_{128}(44,[:,39])$, and $\tilde{B}_{128}(44,[:,47])$. The number of such columns is $Q_2 Q_1$, which is consistent with our previous analysis. Further, to eliminate the impact of $\tilde{B}_{128}(44,[49:64,43])$, the applicable common elements are $\tilde{B}_{128}(44,[:,27])$ and $\tilde{B}_{128}(44,[:,59])$, totaling $Q_2$ elements.

If we want to express columns with fixed indices in the first and second layers, $q_0$ and $q_1$, and varying index in the third layer, $q_2$, the $Q_2$ elements are always required, since the third part of the target columns remains unchanged across different values of $q_2$. Next, if we fix $q_0$ but allow $q_1$ to vary, the second parts of the target columns are all identical. As a result, there are $Q_2 Q_1$ usable elements for removal. Taking Fig. \ref{fig_2_plus_2} as an example, based on relation operations, the specific common elements corresponding to fixed $q_0$ and varying $q_1$ and $q_2$ are expressed as
\begin{align}
q_1=1,q_2=1:&[(2^1_S,2^2_S)^1_S]\label{sequence_example_M_2_1}\\
q_1=2,q_2=1:&[(2^1_S,2^2_S)^2_S]\label{sequence_example_M_2_2}\\
q_1=1,q_2=2:&[(2^2_S,2^2_S)^1_S]\label{sequence_example_M_2_3}\\
q_1=2,q_2=2:&[(2^2_S,2^2_S)^2_S]\label{sequence_example_M_2_4}
\end{align}
which can be summarized by $[(2^{q_2}_S, 2^{Q_2}_S)^{q_1}_S]$. This expression can be further explained in more detail. Suppose the index of the first layer for the target column and the padding column are $q_0=4$ and $q_0=2$, respectively. For $q_1 = q_2 = 1$, the term $2^2_S$ in \eqref{sequence_example_M_2_1} indicates $Q_2$ pairs used to eliminate the influence of $\tilde{B}_{128}(44,[45:48,43])$. Note that the size equals 2 because each pair includes an additional element to remove $\tilde{B}_{128}(44,[65:128,j])$ for $j = 27, 59$.

Next, increase $q_1$ to 2. If we still use the column with indices $q_0=2, q_1=1, q_2=1$ for padding the first part, the common elements remain the same as before. However, if the column used for first-part padding admits $q_0=2, q_1=2, q_2=1$, a different set of $Q_2$ pairs can be applied to remove the third part. In this case, there are $Q_1$ such relations expressed as $(2^1_S, 2^2_S)$, and the corresponding elements differ from each other. As a result, we can apply the SC operation to the structure $(2^1_S, 2^2_S)$, leading to the representation $(2^1_S, 2^2_S)^{q_1}_S$.

Similar reasoning holds for \eqref{sequence_example_M_2_3} and \eqref{sequence_example_M_2_4}. When extending the case from $M = 2$ to $M = 3$, we obtain the new relation as follows,
\begin{align}\label{sequence_example_M_3}
[((2^{q_3}_S,2^{Q_3}_S)^{q_2}_S,4^{Q_2Q_3}_S)^{q_1}_S,((2^{q_3-1}_S,2^{Q_3}_S)^{Q_2-q_2}_S,4^{Q_2Q_3}_S)^{q_1}_S]
\end{align}

Note that the above analysis focuses on $\Omega_k^-$. Due to the nested structure, for a fixed $M$, the number of layer indices and the number of terms in the specific common element relation are $M$ and $2^{M-1}$, respectively. To express such relations concisely, we define a function $\phi_k(\cdot)$, where the arguments correspond to the varying layer indices. For example, we can write $((2^{q_3}_S, 2^{Q_3}_S)^{q_2}_S, 4^{Q_2Q_3}_S)^{q_1}_S$ compactly as $\phi_3(q_1, q_2, q_3)$, and thus, the relation in \eqref{sequence_example_M_3} can be rewritten as $[\phi_3(q_1, q_2, q_3), \phi_3(q_1, Q_2 - q_2, q_3 - 1)]$. The construction of the sequence $E_k^-$ follows an iterative procedure. For $k = 1$, we initialize $E_1^- = (q_M)$, which corresponds to the specific common element relation when $M = 1$. When increasing $M$ to 2 while fixing $q_M$, the term related to $q_M$ remains unchanged for different values of $q_{M-1}$. Therefore, $E_2^-$ is constructed by concatenating the new layer index $q_{M-1}$ with the previous sequence $E_1^-$.

In addition to $E_k^-$, the sequence $E_k^+$ can be derived straightforwardly. When expressing a column with a fixed index $\hat{q}_0$, padding the first part using columns with $q_0 > \hat{q}_0$ or $q_0 < \hat{q}_0$ leads to different configurations. This is because for a column corresponding to index $\mathcal{D}_i(k)$, the indices $\mathcal{D}_i(h)$ for $h > k$ are not applicable. According to this rule, $E_k^+$ can be derived with minor modifications. This layer-dependent behavior explains why the number of elements in $E_k^-$ doubles at each step, which is consistent with the observation that the first term in \eqref{sequence_example_M_3} contains $q_3$, while the second term uses $q_3 - 1$ as its superscript.

The preceding analysis assumes that the first part of the target column is padded using columns within the first layer. For the case where padding occurs at the $k$-th layer with $k > 1$, the first $k$ parts are merged, while the remaining procedure remains unchanged. As a result, the sequences $E_k^-$ and $E_k^+$ can still be derived accordingly.

In conclusion, based on the above discussions, the iterative construction of the sequences ${E_k^-}$ and ${E_k^+}$, as well as the corresponding sets $\Omega_k^-$ and $\Omega_k^+$, can be systematically obtained.

\textbf{Step 7: Analysis of $\Theta_k$.}

With the previous analysis, the derivation of $\Theta_k$ becomes more straightforward. For $V_k^-$ and $V_k^+$, they represent expressions in terms of other columns at the $k$-th layer. The term $(T_{k+1} - T_{k+2})$ contains the common elements for all representations at the $k$-th layer. These elements correspond to those used for padding various parts and removing the influence of previously chosen common elements. For instance, $\tilde{B}_{128}(44,[:,45])$, $\tilde{B}_{128}(44,[:,57])$, and $\tilde{B}_{128}(44,[:,61])$ in Fig. \ref{fig_2_plus_1} are part of the $(T_{k+1} - T_{k+2})$ set.

Finally, it is important to note that the representation size at the $M$-th layer should be $(2^{\varepsilon_L(i)} - 1)$. This is because, at the outermost layer, the representation should contain the minimum number of elements, which aligns with the definition of $\varepsilon_L(\cdot)$.
\end{IEEEproof}

\textbf{Remark 7:} The layer number $M$ plays a crucial role in the algorithm design, as we will demonstrate that it is related to the computational complexity. The value of $M$ increases when there are more layers needed to express a fixed column, which is influenced by a larger $H_L(i)/C_L(i)$. This is because all the `1's can be considered in a single layer if they are adjacent. The "discreteness" arises from the third row of the generation matrix $B_4$, and therefore, we focus on the range corresponding to the Kronecker product between the polarization kernel and the third row of $B_4$. Intuitively, $M$ becomes larger as $L$ continues to increase. Moreover, we mainly consider subchannels where $i \in [1, L/2]$, since there are repetitions when $i > L/2$. In this case, expressing with $i > L/2$ can be treated as the case where $i \in [1, L/2]$. The maximum value of $M$ among all subchannels for a fixed $L$ occurs in the range $[L/4 + 1, 3L/8]$. We then choose the submatrix $\tilde{B}_{L}(i, [L/4 + 1 : 3L/8, L/4 + 1 : 3L/8])$ as the new matrix. We repeatedly focus on the subchannel index from the median to the upper quartile. With this analysis, it can be concluded that
\begin{align}\label{relation_M_L}
M=\Big\lfloor\frac{1}{2}\log \frac{L}{2}\Big\rfloor
\end{align}

The theorem \ref{theorem_1} thoroughly derive the expression in closed form for $R(x_{\mathcal{D}_i(j)}(\mathcal{C}_L(i))|x_{\{\mathcal{D}^c_i,\mathcal{D}_i(1:j-1)\}}(\mathcal{C}_L(i)))$ where $i\bmod4=0$ and $1\leq i\leq L/2$. In the following, several corollaries are provided to discuss the expressions for the remaining 15 cases. Since some of these cases are directly related to the expression in theorem \ref{theorem_1}, we omit the detailed proofs for brevity.

\begin{corollary}\label{corollary_1}
\textit{Let $1\leq i\leq L/2$ and $i\bmod4\neq2$. Then, the expression of $R(x_{\mathcal{D}_i(j)}(\mathcal{C}_L(i))|x_{\{\mathcal{D}^c_i,\mathcal{D}_i(1:j-1)\}}(\mathcal{C}_L(i)))$ is the same as theorem \ref{theorem_1} by substituting $C_L(i)\rightarrow C_L(i+(i\bmod4))$ and
\begin{equation}\label{corollary_1_Q_0}
Q_0=\left\{
\begin{aligned}
&\frac{C_L(i+i\bmod4)}{4},i\bmod4=1\\
&\frac{C_L(i+i\bmod4)}{2},i\bmod4=3
\end{aligned}
\right.
\end{equation}}
\end{corollary}

Corollary \ref{corollary_1} is straightforward, as the target $\tilde{B}_{L}(i,[i:i+(i\bmod4),j])$ can be considered as the first part, and the remaining analysis still holds. Then, we discuss the case $i\bmod4=2$, which is different from corollary \ref{corollary_1}.

\begin{corollary}\label{corollary_6}
\textit{Let $1\leq i\leq L/2$ and $i\bmod4=2$. Then, the expression of $R(x_{\mathcal{D}_i(j)}(\mathcal{C}_L(i))|x_{\{\mathcal{D}^c_i,\mathcal{D}_i(1:j-1)\}}(\mathcal{C}_L(i)))$ is the same as \eqref{theorem_1_final_equation} with the following substitutions,
\begin{equation}\label{corollary_6_Q_0}
Q_0=\frac{C_L(i+2)}{2}
\end{equation}
\begin{align}\label{corollary_6_T_k}
T_k\triangleq 2^{\theta_L(i)-k-1}-1
\end{align}
\begin{align}\label{corollary_6_Theta_0}
\Theta_0=((V_{0}^-,V_{0}^+,\Delta^-,\Delta^+)\leftarrow (T_{1}-T_{2})_S^{\prod_{h=1}^{M}Q_h})
\end{align}
where
\begin{align}\label{corollary_6_V_0_minus}
V_{0}^-\triangleq((T_0)^{\prod_{h=1}^{M}Q_h}_S\leftarrow\Omega_0^-)^{\lfloor\frac{q_0-1}{2}\rfloor}_S
\end{align}
\begin{align}\label{corollary_6_V_0_plus}
V_{0}^+\triangleq((T_0)^{\prod_{h=1}^{M}Q_h}_S\leftarrow\Omega_0^-)^{\frac{Q_0}{2}-1-\lfloor\frac{q_0-1}{2}\rfloor}_S
\end{align}
\begin{align}\label{corollary_6_Delta_minus}
\Delta^-\triangleq((T_{-1})^{\prod_{h=1}^{M}Q_h}_S\leftarrow\Omega_0^-)^{\lfloor\frac{q_0}{2}\rfloor}_S
\end{align}
\begin{align}\label{corollary_6_Delta_plus}
\Delta^+\triangleq((T_{-1})^{\prod_{h=1}^{M}Q_h}_S\leftarrow\Omega_0^-)^{\frac{Q_0}{2}-\lfloor\frac{q_0}{2}\rfloor}_S
\end{align}}

\textit{The remaining definitions and details are the same as the theorem \ref{theorem_1}.}
\end{corollary}

The distinction for $i \bmod 4 = 2$ arises from that when we want to express the column within the first layer, the representation size has two cases, in which the $\tilde{B}_L(i, [i:i+2, j])$ and $\tilde{B}_L(i, [i, j])$ are separately considered as the first part needed to be padded. In this way, the former case has the 'removing-padding' step less 1 than that of the latter one. Moreover, it can be shown that for fixed $q_j, j=1,\cdots, M$, the above two cases keep alternating as the $q_0$ increases, which leads to the flooring operation on the index $q_0$.

Then, we delve into the case $L/2+1 \leq i \leq L$ that is more complicated but similar to theorem \ref{theorem_1}. We provide the complete details but omit the proof. Instead, the difference between $1 \leq i \leq L/2$ and $L/2+1 \leq i \leq L$ are explained, which facilitates the proof based on the analysis of theorem \ref{theorem_1}. Since cases $i \bmod j, j=0, 1, 2, 3$ are also not the same, the $i \bmod 4 = 0$ is first given.

\begin{corollary}\label{corollary_2}
\textit{Let $L/2+1\leq i\leq L$ and $i\bmod4=0$. Define a constant sequence $\{Q_0,\cdots,Q_M\}$ and an index sequence $\{q_0,\cdots,q_M\}$, where $q_j\in\{1,\cdots,Q_j\},j=0,\cdots,M$. Let iterative relation of $t_j$ and $t_{j-1}$ as \eqref{theorem_1_t_j} with $t_0=i\bmod(L/\beta_L(i))$ and $j=1,\cdots,M-1$. The term $Q_j$ is given as
\begin{align}\label{corollary_2_Q}
Q_{M-j+1}=&\min\Big\{\frac{H_L(i)}{C_L(i)\beta_L(i)\prod_{h=1}^{j-1}Q_{M-h+1}},\nonumber\\
&\beta_{2^{\lceil\log t_{j-1}\rceil}}(t_{j-1})\Big\},j=1,\cdots,M
\end{align}
and $Q_0=C_L(i)$. The integer $M$ is the largest index such that $Q_{j},j=0,\cdots,M+1$ are all larger than 1 and we set $Q_{M+1}=\beta_L(i)$. Define
\begin{align}
\phi_k(g,q_{M-k+1}^M)\triangleq&[((\phi_{k-1}(g,q_{M-k+2}^M)),\nonumber\\
&((T_{M-k+3})^{\beta_L(i)}_P)^{\prod_{h=0}^{k-2}Q_{M-h}}_S)^{q_{M-k+1}}_S],\nonumber\\
&\hspace{2.5cm} k=2,\cdots,M
\end{align}
with $\phi_0(g)=((1)^g_K)$ and $\phi_1(g,q_M)=((1)^g_K,(T_{M+1}-1)^{\beta_L(i)}_P)^{q_M}_S$. Then, define column vector sequences $\{E_h^-\},\{E_h^+\}$ for $h=0,\cdots,M$, initialized as $E_0^-=(g)^\top$ and $E_0^-=(g-1)^\top$. The vectors $\{E_h^-\},\{E_h^+\},h=1,\cdots,k$ can be obtained iteratively,
\begin{align}
E_h^-=&((E_{h-1}^-,q_{M+1-h}\mathbf{1}_{2^{h-1}})^\top,\nonumber\\
&(E_{h-1}^+,(Q_{M+1-h}-q_{M+1-h})\mathbf{1}_{2^{h-1}})^\top)^\top
\end{align}
\begin{align}
E_h^+=&((E_{h-1}^-,(q_{M+1-h}-1)\mathbf{1}_{2^{h-1}})^\top,\nonumber\\
&(E_{h-1}^+,(Q_{M+1-h}-q_{M+1-h}+1)\mathbf{1}_{2^{h-1}})^\top)^\top
\end{align}}

\textit{Then,
\begin{align}
\Omega_k^-=&(\phi(E_{M-k}^-(1)),\cdots,\phi(E_{M-k}^-(2^{M-k})))\\
\Omega_k^+=&(\phi(E_{M-k}^+(1)),\cdots,\phi(E_{M-k}^+(2^{M-k})))
\end{align}}

\textit{Finally, we have
\begin{align}\label{corollary_2_final_equation}
&\sum_{j=1}^{|\mathcal{D}_i|}\lambda(R(x_{\mathcal{D}_i(j)}(\mathcal{C}_L(i))|x_{\{\mathcal{D}^c_i,\mathcal{D}_i(1:j-1)\}}(\mathcal{C}_L(i))))\nonumber\\
=&\sum_{g=1}^{\beta_L(i)}\sum_{q_M=1}^{Q_M}\cdots\sum_{q_1=1}^{Q_1}\sum_{q_0=1}^{C_L(i)}\lambda([(1)^{g-1}_P,(\Delta^-,\Delta^+,\Theta_{M-1})\leftarrow\nonumber\\
&\hspace{3cm}(2^{\varepsilon_{L/\beta_L(i)}(t_0)-1}-1)^{\beta_L(i)}_P]),
\end{align}
where for $k=0,\cdots,M-1$, we define
\begin{equation}\label{corollary_2_Theta}
\Theta_{k}\triangleq((V_{k}^-,V_{k}^+,\Theta_{k-1})\leftarrow ((T_{k+1}-T_{k+2})^{\beta_L(i)}_P)_S^{\prod_{h=k+1}^{M}Q_h})
\end{equation}
\begin{equation}\label{corollary_2_Delta_minus}
\Delta^-\triangleq((2^{\varepsilon_{L/\beta_L(i)}(t_0)-1})^{\beta_L(i)}_P\leftarrow(1)^{g}_P)^{q_M-1}_S
\end{equation}
\begin{equation}\label{corollary_2_Delta_minus}
\Delta^+\triangleq((2^{\varepsilon_{L/\beta_L(i)}(t_0)-1})^{\beta_L(i)}_P\leftarrow(1)^{g-1}_P)^{Q_M-q_M}_S
\end{equation}
\begin{equation}\label{corollary_2_V_k_minus}
V_k^-\triangleq(((T_k)^{\beta_L(i)}_P)^{\prod_{h=k+1}^{M}Q_h}_S\leftarrow\Omega_k^-)^{q_k-1}_S
\end{equation}
\begin{equation}\label{corollary_2_V_k_plus}
V_k^+\triangleq(((T_k)^{\beta_L(i)}_P)^{\prod_{h=k+1}^{M}Q_h}_S\leftarrow\Omega_k^+)^{Q_k-q_k}_S
\end{equation}
with $\Theta_{-1}=(0)$.}
\end{corollary}

The main difference between $1\leq i\leq L/2$ and $L/2+1\leq i\leq L$ is that for the latter case, every $\tilde{B}_{L}(i,[:,j]),j=1,\cdots,L/\beta_L(i)$ occurs $\beta_L(i)$ times due to the repetition. Additionally, there is another higher layer with index $g\in\{1,\cdots,\beta_L(i)\}$ in which the target column $\tilde{B}_{L}(i,[:,j])$ is directly equal to $\tilde{B}_{L}(i,[:,j+L/\beta_L(i)])$. Thus, there is `$(1)^{g-1}_P$' in \eqref{corollary_2_final_equation} and a PC operation for almost all elements.

\begin{corollary}\label{corollary_3}
\textit{Let $L/2+1\leq i\leq L$. Then, with similar manipulations, the expressions of $R(x_{\mathcal{D}_i(j)}(\mathcal{C}_L(i))|x_{\{\mathcal{D}^c_i,\mathcal{D}_i(1:j-1)\}}(\mathcal{C}_L(i)))$ with $i\bmod4=j,j=1,2,3$ are extended based on corollaries \ref{corollary_1} and \ref{corollary_2}.}
\end{corollary}

Next, we analyze cases in which the $U_i$ is used as the usable prior information. There are also similarities with the previous analysis, and we only present the distinction compared to cases where the $U_i$ is unknown.

\begin{corollary}\label{corollary_4}
\textit{Let $1\leq i\leq L/2$ and $i\bmod4=j,j=2,3$. Then, $R(x_{\mathcal{D}_i(j)}(\mathcal{C}_L(i+1))|x_{\{\mathcal{D}^c_i,\mathcal{D}_i(1:j-1)\}}(\mathcal{C}_L(i+1)))$ is the same as theorem \ref{theorem_1} by substituting $Q_0$ as \eqref{corollary_1_Q_0} and $T_k\triangleq2^{\Theta_L(i+1)-k}-1$ and
\begin{equation}
\Theta_{-1}=\left\{
\begin{aligned}
&(T_0)^{H_L(i)}_S,i\bmod4=3\\
&(T_0)^{\frac{H_L(i)}{2}}_S,i\bmod4=2
\end{aligned}
\right.
\end{equation}}

\textit{Let $i\bmod4=1$ and then, the expression is the same as corollary \ref{corollary_2} by setting $Q_0=C_L(i+2)/2$, $T_k\triangleq2^{\Theta_L(i+1)-k-1}-1$ and $\Theta_{-1}=(T_{-1})^{H_L(i)}_S$.}

\textit{Finally, let $i\bmod4=0$ and then, it is the same as $i\bmod4=3$ with setting $\Theta_0=(T_1)^{\prod_{h=1}^{M}Q_h}_S$ and $\Theta_{-1}={0}$.
}
\end{corollary}

For $R(x_{\mathcal{D}_i(j)}(\mathcal{C}_L(i+1))|x_{\{\mathcal{D}^c_i,\mathcal{D}_i(1:j-1)\}}(\mathcal{C}_L(i+1)))$ with $1\leq i\leq L/2$, the uncertainty brought by `1's in the $i$-th row has been eliminated. Consequently, additional prior columns become available for utilization, leading to the distinction represented by $\Theta_{-1}$. Specifically, for the element indexed by $\mathcal{D}_i(1)$, its relationship is characterized by $\Theta_{-1}\neq(0)$.

\begin{corollary}\label{corollary_5}
\textit{Let $L/2+1\leq i\leq L$. Then, the expression of $R(x_{\mathcal{D}_i(j)}(\mathcal{C}_L(i+1))|x_{\{\mathcal{D}^c_i,\mathcal{D}_i(1:j-1)\}}(\mathcal{C}_L(i+1)))$ can be extended from $1\leq i\leq L/2$ as $R(x_{\mathcal{D}_i(j)}(\mathcal{C}_L(i))|x_{\{\mathcal{D}^c_i,\mathcal{D}_i(1:j-1)\}}(\mathcal{C}_L(i)))$.}
\end{corollary}

\textbf{Remark 8:} Note that the case where $U_i$ is available with $i=L$ cannot be analyzed based on the previous derivations. The reason is that $I(Y_1^L,U_1^{L-1};U_L)=I(Y_1^L(\mathcal{C}_L(L));U_L)$ and it can be observed that $\mathcal{D}_i^c=\emptyset$. In this scenario, the right-hand side (RHS) of \eqref{transform of MI_2} equals 0. However, the simplified form of $I(Y_1^L, U_1^{L-1}; U_L)$ can be easily derived, namely, $I(Y_1^L(\mathcal{C}_L(L));U_L)=h(Y_1^L(\mathcal{C}_L(L)))-h(Y_1^L(\mathcal{C}_L(L))|U_L)=h(Y)+\sum_{p=1}^{L-1}h_P(p)-Lh(N)$.

\section{$n$-ary tree and pruning algorithm}\label{section_4}
In the previous section, the explicit expressions for the PF under various block lengths and subchannel indexes were fully specified based on the nested structure. The primary information conveyed by the PF is the relationship between $x$ and $x_1^p$. In the following, we present an approach to efficiently compute the corresponding conditional entropy $h(Y|Y_1^p)$, where $y$ and $y_1^p$ represent the channel outputs corresponding to $x$ and $x_1^p$, respectively.

\subsection{Calculation of $h_S(p)$ and $h_P(p)$}
Before examining the general form of $h(Y|Y_1^p)$, which may exhibit an arbitrary representation relationship, we first focus on calculating $h_S(p)$ and $h_P(p)$, which represent special cases of $h(Y|Y_1^p)$ and are simpler to analyze. Compared to $h_S(p)$ and $h_P(p)$ for $p \geq 2$, the analytical expressions for $h_S(1)$ and $h_P(1)$ are much easier to derive. For simplicity, we subsequently consider only $h_S(1)$, as $h_S(1) = h_P(1)$. Based on definition \ref{definition_1},
\begin{align}\label{h_S_p_1_simplification}
h_S(1)=&h(Y,Y_1)-h(Y_1)\nonumber\\
=&h_Y(d=\sqrt{2}A)+h(N)-h_Y(d=A)
\end{align}
where we assume that $X, X_1 \in \{0, A\}$ and denote $h(Y) \triangleq h_Y(d = A)$. Therefore, our approach is to design the merging operation that transforms the calculation of $h_S(p)$ and $h_P(p)$ for $p \geq 2$ into the computation of $h_S(1)$. The closed-form expression for $h_Y(d = A)$ will be analyzed at the end of this subsection.

Next, we set $p \geq 2$ and first consider $h_S(p)$. Before proceeding, we explain the physical mechanism behind $W(y | y_1)$. The correlation between $y$ and $y_1$ arises from the fact that $x = x_1$. Thus, for a given $y_1$, $x_1$ is detected with an error probability $e_1$. The estimated $\hat{x}_1$ is then used as the channel input $x$ to produce $y$. For the SC entropy, $W(y | y_1^p)$ can be expressed as follows:
\begin{align}\label{merging_SC}
W(y|y_1^p)=&\sum_{x\in\mathcal{X}}W(y|x)W(x|y_1^p)\nonumber\\
=&\sum_{x\in\mathcal{X}}W(y|x)\sum_{\bigoplus_{j=1}^{p}x_j=x}W\Big(\bigoplus_{j=1}^{p}x_j|y_1^p\Big)
\end{align}

Thus, the physical mechanism of $W(y | y_1^p)$ can be analyzed similarly. For instance, to obtain the estimate $\hat{x}$ and generate the corresponding $y$ through the underlying channel, each $\hat{x}_j$ must first be inferred from the observed $y_j$ with an error probability of $e_1$. It is important to note that all $Y_j$ are mutually independent and generated by the same underlying channel. As a result, the total error probability that $\hat{x} \neq x$ is given by
\begin{align}\label{combination_detection_error}
e_p=\mathcal{T}(e_1,p)\triangleq\sum_{j=0}^{\lceil\frac{p}{2}\rceil-1}\binom{p}{2j-1}e_1^{2j-1}(1-e_1)^{p-2j+1}
\end{align}

In this way, $W(y | y_1^p)$ becomes equivalent to $W(y | \breve{y}_1)$, where the detection error probability of $\breve{x}_1$ given $\breve{y}_1$ is $e_p$, and $x = \breve{x}_1$. Let $\mathcal{P}(\cdot)$ denote the detection error function that maps the Euclidean distance to the bit error rate (BER). Furthermore, we assume $e_1 = \mathcal{P}(A)$, where $X \in {0, A}$. Under this assumption, the effective Euclidean distance of $\breve{x}_1$ after merging is transformed according to $A \rightarrow \mathcal{P}^{-1}(\mathcal{T}(\mathcal{P}(A), p))$. We refer to the process of consolidating multiple channel outputs $y_1^p$ into a single virtual channel output $\breve{y}_1$ as the `\textbf{merging operation}'. For the PC entropy, the estimation of $\hat{x}$ is equivalent to decoding a repetition codeword of block length $p$, where the corresponding Hamming distance translates to a Euclidean distance of $\sqrt{p}A$. Therefore, the Euclidean distance transformation under the merging operation for PC entropy is $A \rightarrow \sqrt{p}A$.

Combining the above analysis, $h_S(p)$ and $h_P(p)$ can be expressed as follows:
\begin{align}\label{h_S_caculation}
h_S(p)=h_Y(d=\mathcal{P}^{-1}&(\mathcal{T}(\mathcal{P}(A),p)))\nonumber\\
&+h(N)-h_Y(d=A)
\end{align}
\begin{equation}\label{h_P_caculation}
h_P(p)=h_Y(d=\sqrt{p}A)+h(N)-h_Y(d=A)
\end{equation}

Finally, we derive the analytical expression for $h_Y(d = A)$. Due to the presence of the logarithmic function, the channel output entropy is not straightforward to express in closed form. Fortunately, by partitioning the integration domain and applying a polynomial approximation, a closed-form expression for $h_Y(d = A)$ can be analytically obtained. First, we have
\begin{align}\label{expression_h_Y_1}
h_Y(d=A)=&-\int_{-\infty}^{-\frac{A}{2}}W(y)\log W(y)dy-\int_{-\frac{A}{2}}^{+\infty}W(y)\nonumber\\
&\times\log W(y)dy
\end{align}

For the second term of \eqref{expression_h_Y_1},
\begin{align}
&\int_{-\frac{A}{2}}^{+\infty}W(y)\log W(y)dy\nonumber\\
=&\frac{1}{2}\int_{-\frac{A}{2}}^{+\infty}(W(y|0)+W(y|A))\log\frac{(W(y|0)+W(y|A))}{2}dy\nonumber\\
=&\frac{1}{2}\int_{-\frac{A}{2}}^{+\infty}(f_N(n)+f_N(n+A))\log\frac{(f_N(n)+f_N(n+A))}{2}dn\nonumber\\
=&\frac{1}{2}\int_{-\frac{A}{2}}^{+\infty}(f_N(n)+f_N(n+A))\log f_N(n)\nonumber\\
&+\log\Big(1+\frac{f_N(n+A)}{f_N(n)}\Big)dn-1
\end{align}
where $f_N(\cdot)$ denotes the probability density function (PDF) of the channel noise. For $n \in [-A/2, +\infty)$, it holds that $f_N(n) \geq f_N(n + A)$. Consequently, the term $\log(1 + f_N(n + A)/f_N(n))$ can be approximated using a polynomial technique, since the logarithmic function is nearly linear over the interval $[1, 2]$. This allows a low-order polynomial to achieve sufficient approximation accuracy without introducing significant computational complexity. The resulting simplified integral greatly facilitates the derivation of closed-form expressions.

This approximation is central to the design of the efficient polarization decomposition algorithm. For practical purposes, we provide in the following example the conditional entropy expressions and their corresponding fitting results under the assumption of WGN.
\begin{example}\label{example_3}
\textit{Let the channel noise random variable $N$ follow a Gaussian distribution with zero mean and variance $\sigma^2$. Then, the approximated channel output entropy $\hat{h}_Y(d = A)$ can be expressed as follows,
\begin{align}\label{channel_output_entropy_Gaussian}
&\hat{h}_Y(d=A)\nonumber\\
=&\sum_{j=0}^{\rho-1}\gamma_j\exp\Big(\frac{j^2A^2-jA^2}{2\sigma^2}\Big)\sqrt{\frac{\pi}{2}}\sigma\Big[\psi\Big(\frac{-2jA+A}{2\sqrt{2}\sigma}\Big)+1\Big]\nonumber\\
&+\sum_{j=0}^{\rho-1}\gamma_j\exp\Big(\frac{(j+1)^2A^2-(j+1)A^2}{2\sigma^2}\Big)\sqrt{\frac{\pi}{2}}\sigma\nonumber\\
&\times\Big[\psi\Big(\frac{-2(j+1)A-A}{2\sqrt{2}\sigma}\Big)+1\Big]
\end{align}
where $\rho$ is the polynomial order and $\psi(x)\triangleq\frac{2}{\sqrt{\pi}}\int_{0}^{x}\exp(-t^2)dt$. The $\gamma_j$ with $j=0,\cdots,\rho-1$ are the polynomial coefficients which are used to fit $\log(1+x),x\in[0,1]$. The fitting simulation results are presented in Fig. \ref{fig_3}.}

\textit{
\begin{figure}[htbp]
\centering
\includegraphics[width=3.6in]{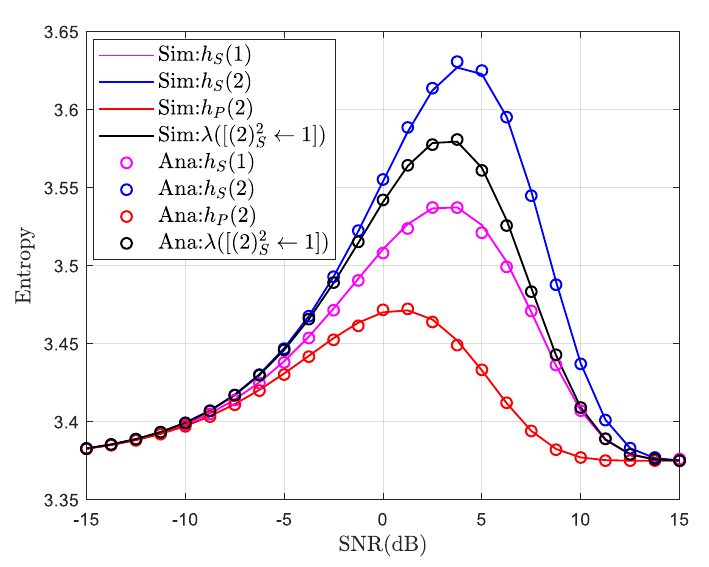}
\caption{The fitting results for the channel output entropy, SC entropy, and PC entropy under WGN are presented. We set that $\rho = 7$. In the legend, `Sim' denotes the entropy values obtained via numerical integration, while `Ana' refers to those derived from analytical expressions.}
\label{fig_3}
\end{figure}
}

\textit{As shown in Fig. \ref{fig_3}, the analytically approximated results closely match the numerically computed values across the entire SNR range. The maximum deviation between the approximation and the numerical integration is approximately 0.005 bits. Moreover, without the proposed analytical method, computing $h_S(p)$ and $h_P(p)$ via numerical integration would be highly time-consuming.}
\end{example}

\subsection{Constructing trees and pruning}
The proposed method for computing $h_S(p)$ and $h_P(p)$ can be extended to evaluate the general conditional entropy under more complex representation relationships. For any PF with a fixed representation relation $R(x, x_1^p)$, we aim to describe this relation using a tree structure, motivated by the nested operations inherent in $R(x, x_1^p)$, as characterized by theorem \ref{theorem_1}. 

Consider the structure $(((a_1)^{b_1}_{S},(a_2)^{b_2}_{S})\leftarrow(a_0)^{b_0}_{S})$, which indicates that there are $(b_1 + b_2)$ total representations of $x$ using $x_1^p$. Among them, $b_1$ representations consist of $a_1$ elements each, and the remaining $b_2$ representations consist of $a_2$ elements each. Additionally, all $b_1 + b_2$ representations share $b_0$ common component pairs, each composed of $a_0$ elements. This structure can be interpreted as a tree, where the component `$(a_0)^{{b_0}}_{S}$' serves as the parent node, and `$(a_1)^{{b_1}}_{S}$' and `$(a_2)^{{b_2}}_{S}$' represent the corresponding child nodes. If further nesting is present in the representation, the tree structure can be recursively extended. In essence, the entire tree is constructed based on the overlapping operations, which are reflected by the edges connecting parent nodes to their respective children.

\textbf{Remark 9:} The maximum degree of any node in the tree corresponds to the number of representation groups with different sizes. For instance, $(((a_1)^{b_1}_{S},\cdots,(a_p)^{b_p}_{S})\leftarrow(a_0)^{b_0}_{S})$ results in a node with $p$ child nodes, where the sizes $a_j$, for $j = 1, \cdots, p$, are mutually distinct. However, the maximum node degree does not impact the overall algorithmic complexity, as different branches of the tree can be pruned in parallel. This is because the elements across distinct child nodes are mutually independent.

\textbf{Remark 10:} The depth of the tree is determined by the number of nesting operations, which is equivalent to the number of layers denoted by $M$. According to theorem \ref{theorem_1}, the overlap operation appears only in $\Theta_k$, $V_k^-$, and $V_k^+$. For each overlap operation within a given $\Theta_k$, the resulting tree has a depth of 2. Each additional level of nesting increases the depth by 1. Therefore, under the structure specified in theorem \ref{theorem_1}, the overall tree depth is $(M+1)$. A similar analysis applies to other scenarios, where the difference in depth is only some constant. It is important to note that the tree depth directly impacts the complexity of the pruning algorithm, as deeper trees generally require more computational resources to process.

\begin{figure}[htbp]
\centering
\includegraphics[width=3.5in]{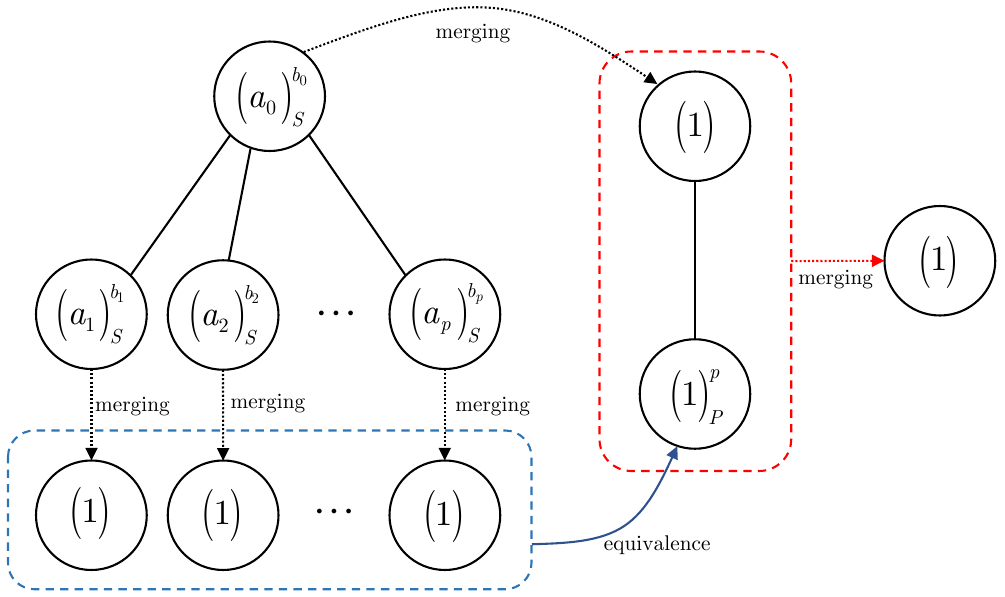}
\caption{The merging process for a two-layer tree. The dotted line is utilized to denote the merging operation. Vertices in a dashed box represent the original elements during a merging operation.}
\label{fig_4}
\end{figure}

Once the tree structure is constructed, we proceed with pruning by repeatedly applying the merging operation for both SC entropy and PC entropy. It is important to emphasize that the primary goal is to compute the conditional entropy $h(Y|Y_1^p)$ for an arbitrary representation relation, with the tree serving to formalize the calculation process. Any tree can be viewed as a combination of nodes and two-layer trees, and Fig. \ref{fig_4} illustrates the pruning procedure.

The relation depicted in Fig. \ref{fig_4} is $(((a_1)^{b_1}_{S},\cdots,(a_p)^{b_p}_{S})\leftarrow(a_0)^{b_0}_{S})$, which corresponds to $h(Y|Y_1^{a_0b_0+\sum_{j=1}^{p}(a_j-a_0)b_j})$. For the first child node `$(a_1)^{b_1}_{S}$', it can be merged to `$(1)$' using the merging operation for SC entropy. This results in $p$ copies of $(1)$. Since the common elements at the higher layer are fixed, these $p$ copies of $(1)$ are identical and can be rewritten as $(1)^p_P$. The merging operation for PC entropy is then applied to convert $(1)^p_P$ into $(1)$. Note that elements corresponding to $(1)$ and $(1)^p_P$ are distinct. Thus, a second merging operation for SC entropy is applied. Finally, the $h(Y|Y_1^{a_0b_0+\sum_{j=1}^{p}(a_j-a_0)b_j})$ is transformed into $h(Y|\breve{Y}_1)$.

Then, we provide a property of merging for SC entropy, which is straightforward and we omit its proof.
\begin{property}\label{property_2}
\textit{Let $R(x,x_1^p)=[(p)]$ and the constant $p_1$ such that $p\bmod p_1=0$. Denote the virtual channel output obtained by merging $Y_1^p$ as $\breve{Y}_1$. Then, the two merging processes, $(p) \to (1)$ and $(p) \to (p_1) \to (1)$, result in the same virtual RV $\breve{Y}_1$.}
\end{property}

During pruning, the SC operation and PC operation can be nested, i.e., $((a_1)^{b_1}_S)^{b_2}_P$. Therefore, the order of merging operations needs to be carefully considered. It can be verified that `PC(SC($\cdot$))' and `SC(PC($\cdot$))' result in $\breve{Y}_1$ with different detection error probabilities. According to \eqref{merging_SC}, for the operation to be valid, the elements $x_j$ with $j=1, \dots, p$ must be distinct. Otherwise, the synthesized detection error cannot be derived using \eqref{combination_detection_error}. In fact, if the elements $x_j$ are not mutually different, the error probability $e_p$ becomes much more complicated, which increases the complexity of the pruning process. Thus, for $((a_1)^{b_1}_S)^{b_2}_P$, the merging for PC entropy should be first applied. This simplifies $((a_1)^{b_1}_S)^{b_2}_P$ to $(a_1)^{b_1}_S$. This analysis demonstrates that the merging for $h_P(p)$ takes precedence over that for $h_S(p)$.

\begin{figure}[htbp]
\centering
\includegraphics[width=3.6in]{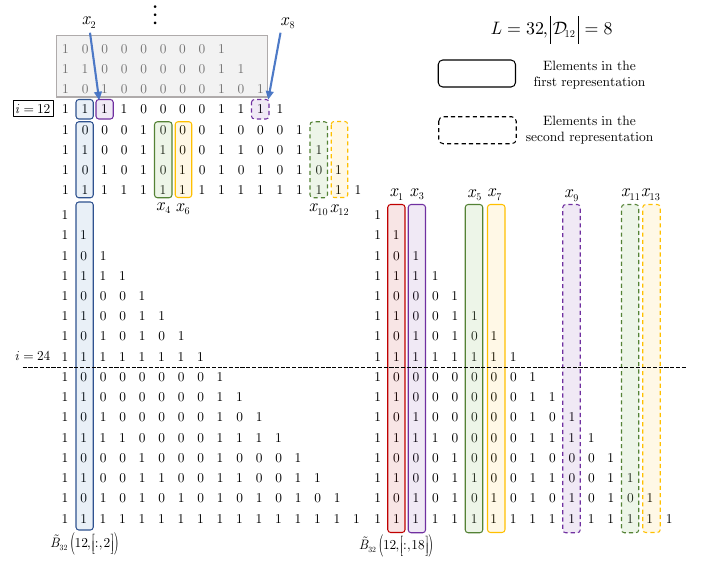}
\caption{The case that the merging order needs to be considered. We set $L=32,i=12$ and aim to represent $\tilde{B}_{32}(12,[:,2])$. We focus on two representations which separately contains $\tilde{B}_{32}(12,[:,3])$ and $\tilde{B}_{32}(12,[:,11])$. The elements within the boxes with solid and dashed lines denote the first and second representations, respectively.}
\label{fig_5}
\end{figure}

The merging order introduces another important issue during pruning: this order must sometimes be considered even when we aim to merge $(a_1)^{b_1}_S$ to $(1)$. To illustrate this phenomenon, consider the example shown in Fig. \ref{fig_5}. We seek to express $\tilde{B}_{32}(12,[:,2])$ based on two representations, depicted by a solid line and a dashed line, respectively. We use $\tilde{B}_{32}(12,[:,3])$ and $\tilde{B}_{32}(12,[:,11])$ to pad the first part. There are a total of 13 elements in the two representations, with the common element $\tilde{B}_{32}(12,[:,18])$ highlighted by the red solid box. The relation is given by $R(x, x_1^{13}) = [(7)^2_S \leftarrow (1)]$, corresponding to a two-layer tree with a parent node $(1)$ and a child node $(6)^2_S$. The common element is denoted as $x_1$. It may initially seem that the merging operation should be the same as in Fig. \ref{fig_4}. For the child node, the merging for SC entropy is applied first. During this process, we have that
\begin{align}\label{example_SC_merging_method_1}
W(x\oplus x_1|y_2^{13})=\sum_{\substack{\bigoplus_{j=2}^{7}x_j=x\oplus x_1\\\bigoplus_{j=8}^{13}x_j=x\oplus x_1}}W(\breve{x}_1,\breve{x}_2|y_2^{13})
\end{align}
where $\breve{x}_1\triangleq\bigoplus_{j=2}^{7}x_j$ and $\breve{x}_2\triangleq\bigoplus_{j=8}^{13}x_j$. The merging can be divided into two steps, as described in \eqref{example_SC_merging_method_2}. In the second merging step, the three elements $(x_2 \oplus x_3, x_4 \oplus x_5, x_6 \oplus x_7)$ are considered. A similar process is applied for $x_8^{13}$. According to property \ref{property_2}, the two merging processes using \eqref{example_SC_merging_method_1} and \eqref{example_SC_merging_method_2} should be equivalent but unfortunately, this is not true. From Fig. \ref{fig_5}, it can be seen that the sum of columns in solid purple boxes is the same as that in dashed purple boxes, as well as the sums in the green and yellow boxes. Specifically, we have $x_2 \oplus x_3 = x_8 \oplus x_9$, $x_4 \oplus x_5 = x_{10} \oplus x_{11}$, and $x_6 \oplus x_7 = x_{12} \oplus x_{13}$. Thus, \eqref{example_SC_merging_method_2} converts $(6)^2_S$ to $(3)^2_P$, not $(3)^2_S$. After applying \eqref{example_SC_merging_method_2}, we must use the merging operation for PC entropy and simplify $(3)^2_P$ to $(3)$, and finally merge $(3)$ to $(1)$. In other words, the merging process in \eqref{example_SC_merging_method_1}, which uses a single SC entropy merging operation, is incorrect.

This example shows that $(a_1)^{b_1}_S$ may not always be merged directly if $a_1$ exceeds a certain threshold $A_t$. For example, if $a_1 = 2$ in Fig. \ref{fig_5}, the merging order is not necessary. The reason is that there are multiple representations with the same size. Even though the elements appear distinct, the sum of a subset of elements in one representation may equal the sum in another representation. If the subset size equals the element number in the representation, $(a_1)^{b_1}_S$ can be simplified directly to $(1)^{b_1}_P$. Otherwise, we first merge the elements in the subset, then apply merging for PC entropy, and finally use the second SC merging to produce $(1)$. In summary, the processes for both cases are as follows,
\begin{equation}
a_1>A_t:(a_1)^{b_1}_S\overset{\text{SC}}{\rightarrow}\Big(\frac{a_1}{A_t}\Big)^{b_1}_P\overset{\text{PC}}{\rightarrow}\Big(\frac{a_1}{A_t}\Big)\overset{\text{SC}}{\rightarrow}(1)
\end{equation}
\begin{equation}
a_1=A_t:(a_1)^{b_1}_S\overset{\text{SC}}{\rightarrow}(1)^{a_1/A_t}_P\overset{\text{PC}}{\rightarrow}(1)
\end{equation}

Note that $a_1\bmod A_t=0$ is always satisfied during the merging process. Based on the previous analysis, it can be straightforward to show that $A_t=L/2^{\lceil\log i\rceil}$ where $i$ is the subchannel index.

\begin{figure*}[hb]
\hrulefill
\begin{align}\label{example_SC_merging_method_2}
W(x\oplus x_1|y_2^{13})=&\sum_{\substack{x_2^5\in\mathcal{X}^4\\x_6\oplus x_7=x\oplus x_1\oplus(\bigoplus_{j=2}^{5}x_j)}}\hspace{0.2cm}\sum_{\substack{x_8^{11}\in\mathcal{X}^4\\x_{12}\oplus x_{13}=x\oplus x_1\oplus(\bigoplus_{j=8}^{11}x_j)}}W((x_2\oplus x_3,x_4\oplus x_5,x_6\oplus x_7),\nonumber\\
&\hspace{8cm}(x_8\oplus x_9,x_{10}\oplus x_{11},x_{12}\oplus x_{13})|y_2^{13})
\end{align}
\end{figure*}

Consequently, given the tree structure, the pruning algorithm can be used to produce the conditional entropy, which is summarized in the algorithm \ref{algorithm_1}. In addition, we are ready to propose the polarization decomposition algorithm, which is provided in algorithm \ref{algorithm_2}.

\begin{algorithm}[!ht]
\renewcommand{\algorithmicrequire}{\textbf{Input:}}
\renewcommand{\algorithmicensure}{\textbf{Output:}}
\caption{Pruning algorithm}\label{algorithm_1}
\begin{algorithmic}[1]
\Require $R(x,x_1^p)$, error probability $e_1$. 
\Ensure Conditional entropy $h(Y|Y_1^p)$. 
\State Denote the depth of the tree as $D_T$.
\State $k=D_T$.
\While {$k>1$}
\State Considering the merging order, perform the corresponding merging operations for all nodes in the $k$-th layer.
\State Merge the nodes in the $k$-th layer and their parent node at the $(k-1)$-th layer via merging for SC entropy.
\State $k=k-1$.
\EndWhile
\State Determine $h(Y|Y_1^p)$ based on \eqref{h_S_caculation} and \eqref{h_P_caculation}.
\end{algorithmic}
\end{algorithm}

\begin{algorithm}[!ht]
\renewcommand{\algorithmicrequire}{\textbf{Input:}}
\renewcommand{\algorithmicensure}{\textbf{Output:}}
\caption{Polarization decomposition algorithm}\label{algorithm_2}
\begin{algorithmic}[1]
\Require Block length $L$. 
\Ensure PF expressions of symmetric capacity of all subchannels $I(W_L^{(i)})$. 
\State $H_L(1)=1$, $\theta_L(1)=\times$, $\theta_L(L)=1$, $K(1:4)=[4,1,2,1]$, $\varepsilon_L(L)=\times$.
\State Determine $H_L(i)$, $\beta_L(i)$, $C_L(i),\theta_L(i)$ and $\varepsilon_L(i)$ for all $1\leq i\leq L$ based on \eqref{iterative_H}$\sim$\eqref{iterative_varepsilon}.
\State Determine representation relations based on theorem \ref{theorem_1} and similar analysis.
\For {$i=1:L$}
\State Calculate all conditional entropies contained in the PFs of $I(W_L^{(i)})$ based on algorithm \ref{algorithm_1}.
\State Obtain $I(W_L^{(i)})$ based on \eqref{transform of MI_3}.
\EndFor
\end{algorithmic}
\end{algorithm}

\textbf{Remark 11:} The accuracy of $I(W_L^{(i)})$ depends solely on the gap between the approximated channel output entropy and the exact values. Furthermore, the accuracy of the $h_Y(d=A)$ approximation is related to the fitting performance of $\log(1+x)$, where $x \in [0,1]$, based on the polynomial technique. Therefore, by increasing the polynomial order, we can reduce the numerical error in $I(W_L^{(i)})$.

\subsection{Complexity analysis}
The complexity of the pruning algorithm primarily depends on the deepest branch of the tree. Based on theorem \ref{theorem_1}, the depth is related to the number of divided layers $M$, as each nesting of $\Theta_k$ corresponds to an increase in the tree branch depth by 1. Moreover, we have $M \leq \frac{1}{2} \log \frac{L}{2}$ from \eqref{relation_M_L}. At each layer, there are no numerical calculations or iterations, as the error probability function and the output entropy expression during the merging operation have already been derived analytically. Additionally, all branch pruning can be performed in parallel. Consequently, the complexity of the pruning algorithm is $\mathcal{O}(\log M) = \mathcal{O}(\log L)$.

For the polarization decomposition algorithm, the complexity arises from two main aspects. First, determining $H_L(\cdot)$, $\beta_L(\cdot)$, $C_L(\cdot)$, $\theta_L(\cdot)$ and $\varepsilon_L(\cdot)$ contributes a complexity of $\mathcal{O}(L)$. The second aspect involves calculating the PF based on relation expressions. For a fixed block length and subchannel index, the expression is fixed and the complexity only depends on the iterative calculation of $E_k^-$ and $E_k^+$, which also admits a complexity of $\mathcal{O}(\log L)$ based on its iterative relation. For the $i$-th subchannel, the number of PF calculations equals the number of `1's in the $i$-th row of $B_L$. Thus, the total number of PFs is the number of `1's in the generation matrix, which can be verified to be $3^{\log L}$. Since the pruning algorithm is applied once for each PF, the complexity of the polarization decomposition algorithm is $\mathcal{O}((L^{\log3} + L) \log L) = \mathcal{O}(L^{\log3} \log L)$.

\section{Applications}\label{section_5}
We have established a framework to characterize the symmetric capacities of polarized subchannels using a general polarization factor. On one hand, this factor exhibits favorable statistical properties, such as convergence and well-defined asymptotic behavior, which simplifies the analysis of relationships among different subchannels and their evolution with respect to the block length. On the other hand, the general PF also enables efficient numerical computation. Consequently, polarization decomposition not only offers theoretical insights but also enhances practical implementations. In the following sections, we elaborate on several key applications.

\subsection{Theoretical insights}
The polarization decomposition algorithm provides a simplified expression for the MI of subchannels, facilitating a clearer analysis of their relationship and the PO. To illustrate this, we examine two representative cases where $L=2$ and $L=4$.
\begin{equation}
I(W_2^{(1)})=h(Y)-h_S(1),\label{L_2_case_1}
\end{equation}
\begin{equation}
I(W_2^{(2)})=h(Y)+h_P(1)-2h(N),\label{L_2_case_2}
\end{equation}
\begin{equation}
I(W_4^{(1)})=h(Y)-h_S(3),\label{L_4_case_1}
\end{equation}
\begin{equation}
I(W_4^{(2)})=h(Y)+h_S(3)-2h_S(1),\label{L_4_case_2}
\end{equation}
\begin{equation}
I(W_4^{(3)})=h(Y)+h_S(1)-h_P(2)-h_P(3),\label{L_4_case_3}
\end{equation}
\begin{equation}
I(W_4^{(4)})=h(Y)+h_P(1)+h_P(2)+h_P(3)-4h(N).\label{L_4_case_4}
\end{equation}

Obviously, \eqref{L_2_case_1}$\sim$\eqref{L_4_case_4} satisfy the relation $2I(W_L^{(i)})=I(W_{2L}^{(2i-1)})+I(W_{2L}^{(2i)})$. Based on the PO of polar code \cite{paper5}, $I(W_L^{(1)})$ is the smallest among $I(W_L^{(i)}),1\leq i\leq L$. Therefore, as the block length increases to infinity, we have $I(W_L^{(1)})\rightarrow0$, which can be directly verified by
\begin{equation}
\lim\limits_{L\rightarrow+\infty}I(W_L^{(1)})=\lim\limits_{L\rightarrow+\infty}h(Y)-h_S(L-1)=0
\end{equation}

The transformed MI also provides an intuitive explanation for the PO. For instance, the inequality $I(W_4^{(3)}) \geq I(W_4^{(1)})$, as required by PO, is validated by equations \eqref{L_4_case_1} and \eqref{L_4_case_3}, e.g., $I(W_4^{(3)})-I(W_4^{(1)})=h_S(1)+h_S(3)-h_P(2)-h_P(3)\geq0$. Meanwhile, the fastest convergence rate among all subchannels can be identified by combining our algorithm with PO. According to PO, $I(W_L^{(1)})$ and $I(W_L^{(L)})$ correspond to the weakest and strongest subchannels, respectively, and hence converge to 0 and 1 at the fastest rate as $L$ increases. On the other hand, based on algorithm \ref{algorithm_2}, we have
\begin{equation}
I(W_L^{(1)})=h(Y)-h_Y(d=\mathcal{P}^{-1}(\mathcal{T}(\mathcal{P}(A),L-1)))
\end{equation}
\begin{equation}
I(W_L^{(L)})=h(Y)+h_Y(d=\sqrt{L}A)-h_Y(d=A)-h(N)
\end{equation}

Therefore, with the aid of the analytical expression of $h_Y(d=A)$, the convergence order can be more easily derived. For instance, based on equation \eqref{channel_output_entropy_Gaussian}, we observe that $I(W_L^{(1)})$ and $I(W_L^{(L)})$ converge to 0 and 1 exponentially with respect to $L$, which may not hold for channel noise governed by heavy-tailed distributions.

\subsection{Polarization visualization}
\begin{figure*}[htbp]
\centering
\subfloat[$\text{SNR}=2\text{dB}$]{\includegraphics[width=3.6in]{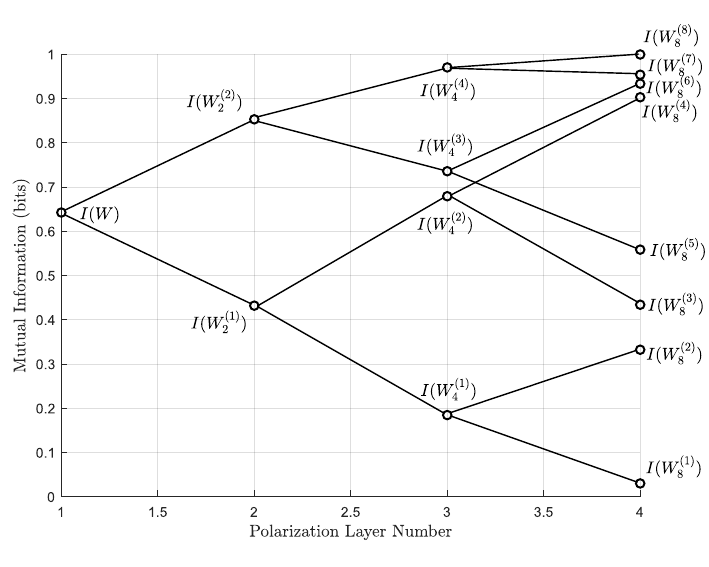}\label{fig_6_1}}
\subfloat[$\text{SNR}=-3\text{dB}$]{\includegraphics[width=3.6in]{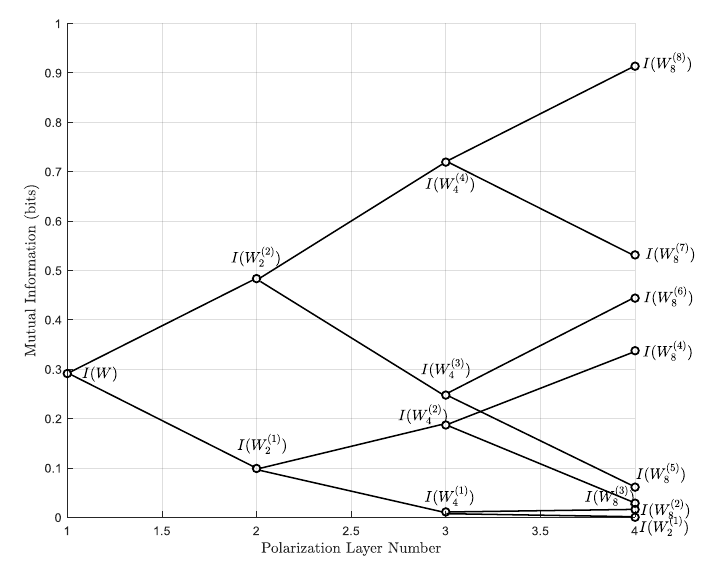}\label{fig_6_2}}
\captionsetup{font=footnotesize}
\caption{The polarization process with block length $L=8$ under WGN and various SNRs.}
\label{fig_6}
\end{figure*}

Based on algorithm \ref{algorithm_2}, we compute the symmetric capacities of subchannels and visualize the polarization process. For a channel with WGN and block length $L = 8$, we evaluate the cases with SNRs of 2dB and -3dB. As shown in Fig. \ref{fig_6}. The lines in the figure represent the fundamental relation $I(W_{2L}^{(2i-1)})+I(W_{2L}^{(2i)})=2I(W_{L}^{(i)})$, while the vertices correspond to the MI values of the underlying channel and polarized subchannels. 

Notably, such comprehensive visualization was previously only tractable for the BEC due to the exponential computational complexity growth with respect to $L$. Implemented on the CPU configuration `12th Gen Intel(R) Core(TM) i9-12900H', \textbf{our algorithm computes all vertices at a fixed SNR in Fig. \ref{fig_6} within approximately 0.0124 seconds, demonstrating a significant speed advantage over conventional numerical integration approaches.}

\subsection{polar code construction}
The core principle of polar code construction lies in selecting the optimal index set $\mathcal{A}(L)$ such that
\begin{equation}
\mathcal{A}(L)=\arg\max\limits_{\mathcal{A}(L)}\sum_{i\in\mathcal{A}(L)}I(W_L^{(i)})
\end{equation}

Existing approaches typically assess subchannel reliability via the Bhattacharyya parameter $Z(W_L^{(i)})$. However, obtaining exact values of $Z(W_L^{(i)})$ remains computationally challenging, necessitating the design of lower and upper bounds. The effectiveness of Bhattacharyya-based construction stems from its fundamental relationship with mutual information, i.e., $Z(W_L^{(i)})^2+Z(W_L^{(i)})^2\leq1,Z(W_L^{(i)})+I(W_L^{(i)})\geq1$ \cite{paper1}. Motivated by this, we propose using mutual information directly as the subchannel quality metric for polar code construction, which aligns more naturally with the theoretical foundations of polar codes.

\begin{figure*}[htbp]
\centering
\subfloat[$L=16$]{\includegraphics[width=3.6in]{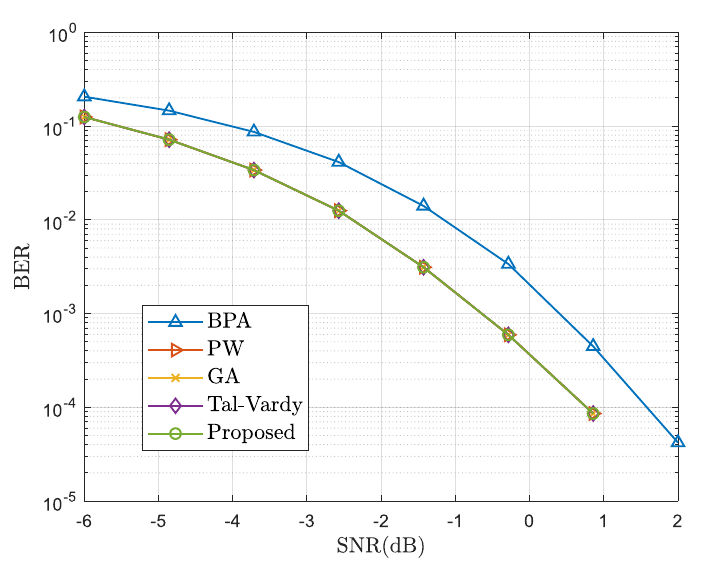}\label{fig_7_1}}
\subfloat[$L=256$]{\includegraphics[width=3.6in]{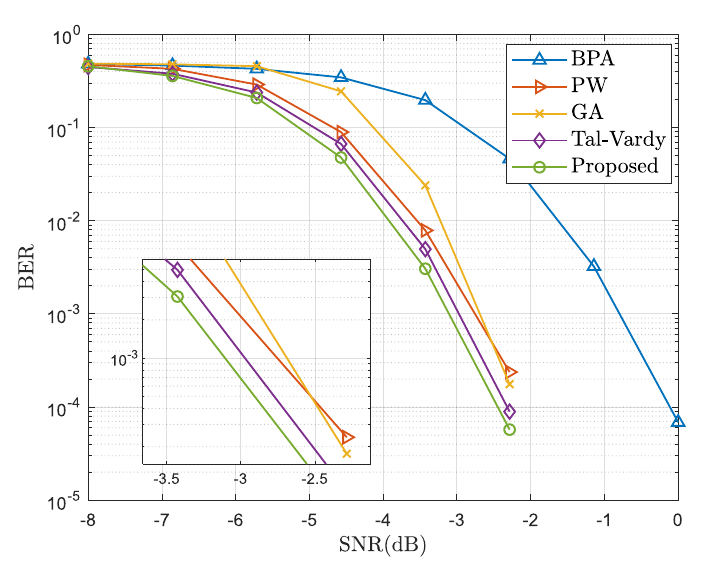}\label{fig_7_2}}
\captionsetup{font=footnotesize}
\caption{The decoding performance comparisons of different polar code algorithms under WGN and various block lengths.}
\label{fig_7}
\end{figure*}

To validate our approach, we present numerical simulations under WGN conditions for block lengths $L=16$ and $L=256$. The code rate is fixed at $R=0.5$, and decoding is performed using the successive cancellation algorithm. We compare against four existing methods including Bhattacharyya parameter approximation (BPA) \cite{paper7}, Gaussian approximation (GA) \cite{paper9}, Tal-Vardy construction \cite{paper8} and polarization weight (PW) construction \cite{paper13}. The performance comparisons are presented in Fig. \ref{fig_7}.

Fig. \ref{fig_7_1} demonstrates nearly identical performance across all methods except BPA. This equivalence occurs because for $L=16$, the selection of $\mathcal{A}(L)$ is relatively straightforward, allowing all competent approaches to identify the optimal subchannels. Consequently, our algorithm shows no performance advantage over the baselines in this case.

As $L$ increases, the selection complexity of $\mathcal{A}(L)$ grows significantly. As shown in Fig. \ref{fig_7_2}, our construction algorithm achieves a 0.2 dB performance gain over the Tal-Vardy method at an SNR of $10^{-3}$. It is worth noting that the BPA method directly applies the Bhattacharyya parameter evolution of the BEC case to arbitrary BMCs, which results in substantial performance degradation. The PW method is empirical in nature, as it involves a hyperparameter that must be tuned through extensive simulations. Furthermore, it is a channel-independent coding scheme, since the construction process does not require any information about the channel, such as the noise distribution or SNR. This lack of adaptability limits the general applicability of the PW algorithm, which is primarily designed for AWGN channels. In contrast, our approach provides universal applicability to all BMCs, with no reliance on channel-specific assumptions during the derivation. The GA algorithm is also limited to AWGN channels, and its piecewise approximation function must be redesigned for other channel types, which is a complex process. The Tal-Vardy method accurately tracks the likelihood ratio evolution through channel degradation and upgrading procedures, thereby maintaining robust baseline performance.

To support our theoretical analysis, we provide an empirical comparison of the computational complexity of various construction methods in Table \ref{table_2}. The results show that our approach achieves computational efficiency through two key innovations. First, we develop a pruning algorithm that reduces the computation of $h(Y|Y_1^p)$ to $h(Y|\breve{Y}_1)$, substantially decreasing the computational cost. Second, by deriving closed-form expressions for both the channel output entropy and PFs, our method eliminates the need for numerical integration and infinite summation.

\begin{table}[htbp]
\begin{center}
\caption{Time consumption of various code construction methods under different block lengths}\label{table_2}
\begin{tabular}{cccccc}
\toprule
$L$&\textbf{Proposed}&Tal-Vardy&GA&PW&BPA\\
\midrule
$16$& \textbf{0.043s}& 0.056s& 0.009s& 0.006s& 0.005s\\
$256$& \textbf{0.739s}& 0.881s& 0.028s& 0.013s& 0.016s\\
\bottomrule
\end{tabular}
\end{center}
\end{table}

\subsection{Rate loss}
Polar codes can asymptotically achieve the channel capacity as the block length increases. However, in practical scenarios, this is not attainable due to the finite value of $L$, which leads to insufficient channel polarization. In such cases, even if $R$ subchannels with the highest capacities are selected to carry information bits, the MI sum of the subchannels indexed by $\mathcal{A}^c(L)$ remains non-zero, resulting in a rate loss. In practical applications, given a specific block length and an information bit index set $\mathcal{A}(L)$, we are interested in quantifying the gap between the achievable rate $\tilde{R}$ and the channel capacity $I(W)$. With the proposed polarization decomposition algorithm, this task can be efficiently addressed, provided that the values of $I(W_L^{(i)}), i = 1, \cdots, L$, are accurately computed. For example,
\begin{equation}
I(W)-\tilde{R}=\frac{1}{L}\sum_{i\in\mathcal{A}^c(L)}I(W_L^{(i)})
\end{equation}

Notably, this capacity-gap analysis arises naturally as a computational byproduct of the construction process, without incurring any additional algorithmic complexity. 

\textbf{In summary, the polarization decomposition approach offers both theoretical insights and practical utility, representing a meaningful contribution to the design and analysis of polar codes.}

\section{discussions}\label{section_6}
\subsection{Conclusion}
This paper investigates the decomposition of the polarization process for BMCs. We reformulate the symmetric capacity of polarized subchannels as a representation problem among the codeword bits within corresponding subcodes. Based on this reformulation, we define the PF and decompose the original MI expression as a combination of PFs. To enable accurate PF computation, we derive its closed-form expression under various scenarios and propose a pruning algorithm along with analytical formulas for channel output entropy. Furthermore, the proposed framework offers both theoretical insights and practical applications, including the analysis of symmetric capacity relationships, encoding structures and rate loss.

\subsection{Future works}

Throughout this work, we assume that the input RVs are i.i.d. and the channel is memoryless. However, in many practical scenarios, the channel or input source may exhibit memory or correlation, which substantially complicates the analysis. For example, the expressions for $h_S(p)$ and $h_P(p)$ become significantly more intricate due to statistical dependencies among channel outputs. Even when $X_1 = X_2$ and $X_1$ is independent of $X_3$, the equality $h(Y_1|Y_2, Y_3) = h(Y_1|Y_2)$ does not hold, as $Y_3$ remains correlated with $Y_1$ through the chain $Y_3 \rightarrow Y_2 \rightarrow X_2 \rightarrow Y_1$. Consequently, the proposed merging processes for $h_S(p)$ and $h_P(p)$ are not valid in such cases. Similarly, the mutual information simplification in \eqref{transform of MI_1} becomes invalid, since channel outputs are mutually dependent when the noise has memory.

Two potential approaches can be considered to address this challenge. The first is to eliminate memory through sufficient interleaving, though the statistical properties of the resulting noise must then be carefully analyzed. The second is to design an approximate decomposition. When noise samples are correlated, the channel outputs become dependent, but this correlation typically decays with increasing time separation \cite{paper14}. Thus, it is possible to truncate the channel output sequence if the correlation falls below a certain threshold. In this case, defining an appropriate correlation metric and analyzing the resulting approximation error are essential.

Meanwhile, with suitable modifications, the proposed framework can be extended to more general settings including $q$-ary input channels, discrete memoryless channels and non-standard polarization processes such as universal polarization and polarization with high-dimensional kernels. The polarization decomposition can also be adapted to other variants of polar codes. For instance, the framework is directly applicable to CRC-aided polar codes, as the concatenated CRC merely serves to validate decoding results. In the case of polarization-adjusted convolutional (PAC) codes, additional modifications are required due to the convolutional transformation. In particular, the impact of the convolutional pre-coding on the input to the polar encoder must be carefully considered.

\begin{appendices}
\section{Proof of property \ref{property_1}}\label{appendix_property_1}
\setcounter{equation}{0}
\renewcommand{\theequation}{A.\arabic{equation}}
For $h_S(p)$, we have
\begin{align}\label{property_equation_0}
W(y|y_1^p)=&\sum_{x\in\mathcal{X}}W(y|x)W(x|y_1^p)\nonumber\\
=&\sum_{x\in\mathcal{X}}W(y|x)\sum_{x=\bigoplus_{j=1}^{p}x_j}W(x_j|y_j)
\end{align}
where $W(y|y_1^p)$ and $W(x|y_1^p)$ denote the corresponding conditional PDF, defined with respect to the underlying channel transition function $W(y|x)$. Denote that $f(0|y_j)\triangleq\alpha_j$ and $f(1|y_j)\triangleq(1-\alpha_j)$ with $x_j=1$. Then,
\begin{align}
W(1|y_1^p)=&\sum_{x=\bigoplus_{j=1}^{p}x_j}(1-x_j)\alpha_j+x_j(1-\alpha_j)\nonumber\\
=&1-\sum_{i\in\{1,\cdots,p\}}\alpha_i+2\sum_{\substack{i\neq j\\i,j\in\{1,\cdots,p\}}}\alpha_i\alpha_j\nonumber\\
&-2^2\sum_{\substack{i\neq j\neq k,i\neq k\\i,j,k\in\{1,\cdots,p\}}}\alpha_i\alpha_j\alpha_k+\cdots
\end{align}

Without loss of generality, we assume that $\alpha_1\leq\alpha_2\leq\cdots\leq\alpha_p$. Define that
\begin{align}
\overline{W}(1|y_1^p)\triangleq&\sum_{x=\bigoplus_{j=1}^{p}x_j}(1-x_j)\alpha_p+x_j(1-\alpha_p)\nonumber\\
\underline{W}(1|y_1^p)\triangleq&\sum_{x=\bigoplus_{j=1}^{p}x_j}(1-x_j)\alpha_1+x_j(1-\alpha_1)
\end{align}

Clearly, we have $\underline{W}(1|y_1^p)\leq W(1|y_1^p)\leq\overline{W}(1|y_1^p)$ since the error probability of detecting $x$ decreases when all $x_j,j = 1, \dots, p$ exhibit the lowest individual detection error $\alpha_1$. As a result, the analysis of $W(1|y_1^p)$ can be reduced to investigating the monotonicity of its upper and lower bounds, $\overline{W}(1|y_1^p)$ and $\underline{W}(1|y_1^p)$, with respect to $p$. We first consider the behavior of $\overline{W}(1|y_1^p)$, which is
\begin{align}\label{property_equation_1}
\overline{W}(1|y_1^p)=&1-\sum_{i=1}^{p}(-2)^{i-1}\binom{p}{i}\alpha_p^i
\end{align}

Based on the following equation
\begin{align}\label{property_equation_2}
(1-2a)^p-1=\sum_{i=1}^{p}(-2)^{i}\binom{p}{i}\alpha_p^i,\nonumber
\end{align}
it follows that for any $0<\alpha_p\leq1/2$, $\lim\limits_{p\rightarrow+\infty}\overline{W}(1|y_1^p)=1/2$. A similar argument applies to $\underline{W}(1|y_1^p)$. Therefore, we obtain $\lim\limits_{p\rightarrow+\infty}W(1|y_1^p)=1/2$. Substituting this result into \eqref{property_equation_0}, we conclude that $\lim\limits_{p\rightarrow+\infty}W(Y|Y_1^p)=W(Y)$, thereby proving \eqref{limit_SC_entropy_order}.

As for $h_P(p)$ defined in \eqref{PC_definition}, based on the Bayesian formula,
\begin{align}
W(\tilde{y}|\tilde{y}_1^p)=&\sum_{x\in\mathcal{X}}W(\tilde{y}|x)W(x|\tilde{y}_1^p)\nonumber\\
=&\sum_{x\in\mathcal{X}}W(\tilde{y}|x)\frac{\prod_{j=1}^{p}W(\tilde{y}_j|x)}{\prod_{j=1}^{p}W(\tilde{y}_j|0)+\prod_{j=1}^{p}W(\tilde{y}_j|1)}
\end{align}

In fact, the underlying input $\tilde{x}_1^p$ corresponding to $\tilde{y}_1^p$ belongs to the set ${\mathbf{0}_p, \mathbf{1}_p}$ and their Euclidean distance in the $p$-dimensional space is $\sqrt{p}$. Assuming that $\tilde{x} = 1$, the observation $\tilde{y}_1^p$ is more likely to be closer to $\mathbf{1}_p$, which makes $\prod_{j=1}^{p} W(\tilde{y}_j | 0)$ significantly smaller than $\prod_{j=1}^{p} W(\tilde{y}_j | 1)$, especially as $p$ becomes large. Consequently, for a given $\tilde{y}_1^p$ and as $p \rightarrow +\infty$, we have $W(0 | \tilde{y}_1^p) \rightarrow 0$ if $\tilde{y}_1^p \in \Xi$, and $W(0 | \tilde{y}_1^p) = 1$ if $\tilde{y}_1^p \in \Xi^c$, where
\begin{align}
\Xi\triangleq\bigg\{\tilde{y}_1^p:\sum_{j=1}^{p}\tilde{y}_j\leq\frac{1}{2}\bigg\}
\end{align}

Hence,
\begin{align}
\lim\limits_{p\rightarrow+\infty}W(\tilde{y}|\tilde{y}_1^p)=W(\tilde{y}|\tilde{x})
\end{align}

\end{appendices}

\footnotesize
\bibliographystyle{IEEEtran}
\bibliography{ref}

\end{document}